\documentclass{jpp}

\usepackage{graphicx}
\usepackage[utf8]{inputenc}
\usepackage[T1]{fontenc}
\usepackage{amsmath}
\usepackage{mathptmx}
\usepackage{etoolbox}
\usepackage{color}
\usepackage{graphicx}
\usepackage[normalem]{ulem}
\usepackage{soul}
\usepackage{dcolumn}
\usepackage{bm}
\usepackage{comment}
\usepackage{amsfonts}
\usepackage{stackengine}
\usepackage{wrapfig}
\usepackage[x11names]{xcolor}
\usepackage{vwcol}
\usepackage{appendix}
\usepackage{hyperref}

\definecolor{darkgreen}{RGB}{20,150,40}

\shorttitle{Helicon waves in toroidal magnetic configurations}
\shortauthor{S. P. H. Vincent et al.}

\title{Helicon wave propagation, plasma generation  \\
and interaction with low-frequency waves \\ in toroidal magnetic configurations}

\author{Simon P. H. Vincent\corresp{\email{sph.vincent@protonmail.com}}, Mounir Alfazzaa, Patrick Quigley, Cyrille Sepulchre, Philippe Guittienne, Rémy Jacquier, Marcelo Baquero-Ruiz, Ivo Furno}

\affiliation{École Polytechnique Fédérale de Lausanne (EPFL), Swiss Plasma Center (SPC), CH-1015 Lausanne, Switzerland}

\begin{document}

\maketitle

\begin{abstract}

Helicon waves are widely used for efficient plasma production in low-temperature devices and have recently attracted attention as a candidate for current drive in fusion plasmas. Yet experimental investigations of helicon waves in toroidal geometries, and of their interaction with plasma dynamics, remain extremely limited. In this work, we present, to our knowledge, the first detailed experimental characterization of helicon waves in a toroidal configuration.
A birdcage resonant antenna operating at 13.56 MHz is used to launch helicon waves in the toroidal basic plasma physics device TORPEX, either into a pre-existing magnetron-generated plasma, or as the plasma source. Measurements are performed in pure toroidal and simple magnetized torus magnetic configurations, for both argon and hydrogen plasmas. Three-axis magnetic probe measurements enable clear identification of a dominant $m=+1$ helicon mode over our parameter space.
The helicon amplitude is found to scale linearly with the antenna power, and decreases with the confining magnetic field amplitude. As the antenna power is increased the helicon amplitude exhibits a saturation, correlated with enhanced low-frequency fluctuations and turbulent transport. In addition, a strong interaction between helicon waves and low-frequency density fluctuations is observed, revealing a non-linear coupling between RF waves and plasma turbulence. 
These results provide the first detailed experimental characterization of helicon waves in a toroidal low-temperature plasma device and establish TORPEX as a unique testbed for studying toroidal helicon wave physics under controlled and well-diagnosed conditions.

\end{abstract}

\newpage

\section{Introduction}

Helicon waves, also referred to as whistlers, or fast waves in the lower hybrid range of frequency, belong to the right-hand polarized branch of electromagnetic waves propagating in a cold plasma immersed in a magnetic field, with a frequency $ \omega_{ci} \ll \omega \ll \omega_{ce}$.

Helicon waves have attracted since the 1970s a lot of attention in the low-temperature plasma (LTP) community for their plasma generation capabilities. It was discovered at the time that inductively coupled plasma sources (ICP) that were specifically designed to excite helicon waves could generate, in the plasma core, relatively higher plasma densities than conventional ICP sources. Since then, extensive theoretical and experimental work has been devoted to helicon waves~\citep{Boswell_1984, Chen_1997, Canases_2016, Filleul_2023, Guittienne_2025}, and various designs of the so-called helicon sources have been developed, that are now routinely used in a large number of plasma laboratory devices around the world. These studies of helicon waves in the context of low-temperature plasma generation has mostly been restricted to linear geometry so far. Yet, the use of helicon sources is not limited to linear devices. Helicon sources are for instance explored for plasma thrusters, with newly investigated designs that include a curved and even a toroidal geometry~\citep{Merino_2023, Shumeiko_2025}. They are also explored as an alternative method for more efficient wall-conditioning using LTP discharges in tokamaks, for example in the EAST tokamak~\citep{Huang_2020}. However, with the exception of a few table-top toroidal devices~\citep{Zhang_1995, Sakawa_2004, Paul_2009}, the fundamental characteristics of toroidal helicon waves in LTP discharges remain largely unexplored.

More recently, helicon waves have gained interest in the context of nuclear fusion, as they are considered as a promising candidate for current drive in tokamaks~\citep{Pinsker_2015}. In this case, helicon waves are not considered for the plasma generation capabilities like in LTP, but for their frequency range that grants them good accessibility to the high-density fusion plasma core.
High-efficiency current drive techniques, with good accessibility close to the plasma core, are indeed essential to achieve steady-state fusion plasmas in tokamaks~\citep{Gormezano_2007}.
The most common techniques using RF waves for current drive include electron cyclotron current drive (ECCD), ion cyclotron current drive (ICCD), and lower hybrid current drive (LHCD). \footnote{ECCD has the advantage of accessibility to high plasma density regions and very accurate control of the wave absorption location, leading to many applications such as q-profile shaping and MHD instability mitigation.
However, ECCD is usually not the best option for non-inductive current drive, and powerful EC launchers are more efficiently used for plasma heating~\citep{Prater_2004}. ICCD, using the fast wave in the ion cyclotron range of frequency, mainly accelerates ions in the parallel direction through Landau damping, which is also insufficient for current drive generation in reactor-grade plasmas. IC wave antennae are therefore likewise mainly used for plasma heating, and for current drive as a complementary source~\citep{Dumont_2013}. In contrast, LHCD, which uses the slow waves in the hybrid range of frequencies to accelerate fast electrons in the parallel direction, has proven to be the most efficient current drive technique to date~\citep{Bonoli_2014, Fonghetti_2025}. But due to the lower-hybrid resonance this slow wave propagation has a density limit, giving LHCD low accessibility to the fusion plasma core in most tokamaks~\citep{Hooke_1984}.  
Helicon waves, fast waves within the lower hybrid range of frequency, do not suffer from a density limit at the LH resonance.}
Theoretical analysis and numerical simulations have predicted that in high-density plasmas such as the ones expected in ITER or DEMO, helicon current drive (HCD) could lead to current drive efficiencies similar to LHCD, but with better access to the plasma core~\citep{Vdovin_2013, Prater_2014}.
In recent years, planar helicon antennae have been mounted on the DIII-D and KSTAR tokamaks~\citep{VanCompernolle_2021, Wi_2023}. The results of the recent first experimental campaigns in DIII-D are promising, with more than 0.7~MW of RF power coupled from the helicon antenna to the plasma~\citep{Pinsker_2024}. HCD is now progressively being considered for a few other tokamaks, and many numerical studies have been dedicated to predicting helicon propagation in a tokamak plasma and optimizing the wave energy deposition~\citep{Li_2020, Li_X_2020, Wu_2023, Gao_2026}. These studies use ray tracing codes, considering a steady-state plasma and focusing on the helicon wave itself.
However, low-frequency plasma waves, plasma turbulence, and fast particles are known to affect RF wave propagation~\citep{Snicker_2018, Chellai_2018}. On the other hand, RF waves impact many aspects of the plasma equilibrium and dynamics, beyond just current drive~\citep{Peysson_2011, Cazabonne_2023}.
In the case of helicon waves, it is worth mentioning a couple of recent studies focusing on the impact of density fluctuations on helicon waves~\citep{Kim_2026} and the effect of helicon waves on run-away electrons~\citep{Choudhury_2025}.
Nonetheless, the effects of the helicon wave interaction with fusion plasma phenomena remain largely unexplored and challenging to study experimentally in a tokamak.

While complex phenomena are difficult to accurately measure and assess in fusion plasmas, experimental studies in well-diagnosed and simplified plasma conditions can prove to be insightful. As such, low-temperature plasma (LTP) devices have often proven to be an ideal tool to shed light on complex fusion-relevant phenomena, from the initial discovery of drift-wave-induced anomalous turbulence to some more recent fast ion transport properties, among others~\citep{Hoh_1962, Fasoli_2019}.
Following this line of work, the recent surge of interest in HCD for future tokamaks calls for the study of toroidal helicon waves on a fundamental and experimental level.

This motivated the design and installation of a helicon antenna on TORPEX, a toroidal low-temperature plasma device of major radius 1 m and minor radius 0.2 m~\citep{Vincent_2024}.
With a power of up to 7~kW, the helicon antenna is able to produce a few $10^{17}$~m$^{-3}$, and helicon waves were successfully measured across the plasma volume in preliminary measurements~\citep{Vincent_2024}.
With its extensive set of diagnostics and its capabilities for code benchmarking~\citep{Ricci_2009}, TORPEX is the ideal testbench for the fundamental and experimental investigation, in a toroidal geometry, of helicon wave properties and their interaction with plasma turbulence and fast particles~\citep{Bovet_2015, Sepulchre_2025}.
We note that the ability of the helicon source to drive current in TORPEX will not explored here. Preliminary measurements indicated that TORPEX regimes may not be relevant for such additional studies.

In this work, we present the results of an extensive experimental campaign that was aimed at characterizing helicon wave propagation in TORPEX, exploring the dependence on the gas species, base pressure, magnetic field, source power, and magnetic field configuration. 
This article constitutes a first experimental step towards the understanding of toroidal helicon wave properties.

We start by introducing in Sec.~\ref{sec::experimental_setup} the experimental setup and diagnostics used in this work.
In Sec.~\ref{sec::helicon_identification}, a simplified model is compared to experimental measurements to unambiguously identify the dominant helicon mode launched in TORPEX.
Helicon wave propagation on top of a background magnetron-generated plasma is investigated in Sec.~\ref{sec::magnetron_generated_plasma}, where the influence of the magnetic configuration and the interaction with low-frequency plasma waves are discussed.
The ability of the helicon antenna to generate plasma and the impact of the antenna power on both plasma parameters and helicon wave properties are then examined in Sec.~\ref{sec::power_scan}. The impact of the magnetic field configuration and strength is explored in Sec.~\ref{sec::mag_field_scans}.
Finally, the main results are summarized and discussed in Sec.~\ref{sec::conclusion}, together with perspectives for future studies on helicon wave interaction with plasma turbulence and fast particles in toroidal geometries.

\section{Experimental setup}
\label{sec::experimental_setup}

\subsection{TORPEX device and magnetic configurations}

\begin{figure}
    \centering
    \includegraphics[width = 0.95\columnwidth, trim={0in 0in 0in 0in},clip]{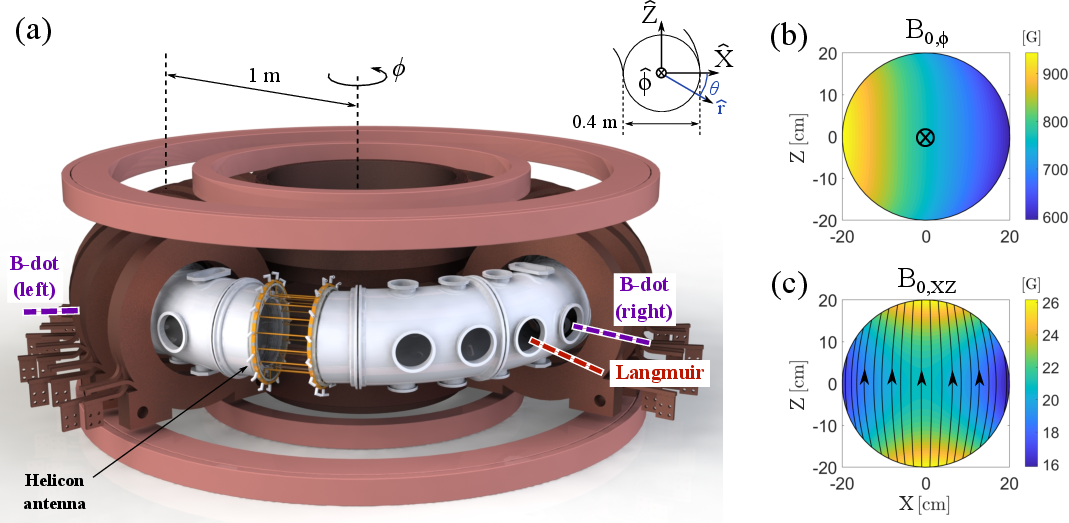}
    \caption{a) Schematic of TORPEX. The Langmuir probe (red dashed line) is inserted at the toroidal location $\phi \simeq + 80 \, ^{\circ}$, and the B-dot probes (purple dashed lines) at the locations $\phi \simeq -80 \, ^{\circ}$ and $\phi \simeq + 100 \, ^{\circ}$, on the left and right side of the helicon antenna respectively. Note that toroidal coils on TORPEX fully surround the chamber along the toroidal direction: a set of these coils is not shown in the figure for a better visualization of the chamber. b) Toroidal magnetic field amplitude. c) Vertical magnetic field amplitude added in the SMT (or "helimak") configuration.}
    \label{torpex_sketch}
\end{figure}

TORPEX is a low-temperature toroidal plasma device of major and minor radii $1$~m and $20$~cm respectively~\citep{Fasoli_2006}. The stainless steel chamber of TORPEX is surrounded by a set of 28 toroidal coils and 5 pairs of poloidal coils, enabling a wide variety of magnetic configurations, with values of the total magnetic field up to $1000$~G. 
Hydrogen or argon plasmas can be generated and maintained by a 2.45~GHz magnetron at gas fill pressures of $\sim 10^{-4}$~mbar up to $\sim 10^{-3}$~mbar.
The main vessel of TORPEX is composed of 12 toroidal sectors (see figure~\ref{torpex_sketch}). One of these sectors is made of two flanges and a borosilicate tube surrounded by a birdcage resonant helicon antenna~\citep{book_Guittienne}, whose technical details are described in \citep{Vincent_2024}. We use a simple toroidal coordinate system $(\hat{X}, \hat{\phi}, \hat{Z})$ as shown in figure~\ref{torpex_sketch}, with the origin $(0, 0, 0)$ taken at the center of the antenna. The antenna is fed by a 13.56 MHz RF generator and is able to launch $m=1$ helicon waves in a pre-existing magnetron plasma, as well as to generate plasma with a RF power up to $7 $~kW.

Two magnetic configurations are used in this work. 
The first one is composed of a pure toroidal field, whose magnitude across a poloidal section of TORPEX is shown in figure~\ref{torpex_sketch}(b), for an average field value of $B_0 = 780$~G. This configuration is of interest in the context of helicon fundamental studies due to its simplicity. However, as is well known in the fusion community, a pure toroidal (PT) magnetic configuration leads to charge separation in the plasma, causing strong drifts and poor plasma confinement. The simple addition of a small vertical field limits this charge separation and significantly improves plasma confinement~\citep{Muller_2004}.
The second configuration that is used in this article consists in a dominant toroidal field with the addition of a small vertical field, as shown in figure~\ref{torpex_sketch}(c).
We refer to this configuration as the Simple Magnetized Torus (SMT)~\citep{Ricci_2010}. Note that elsewhere it has been called the helimak configuration~\citep{Zimmerman_1993}. 
In this work, we choose an SMT configuration with a ratio of the average poloidal field to the average toroidal field $B_{0,(X,Z)} / B_{0,\phi}$ of $\approx 2.5 \%$, which was shown to maximize the confinement in TORPEX~\citep{Muller_2004}.   
In the following, the toroidal and vertical field strengths are referred to by their average value across a poloidal section.

\subsection{Probes}

The plasma density $n$ and floating potential $V_f$ are measured with a Langmuir probe, scanning radial positions at $Z=0$~cm and at a toroidal angle $\phi \simeq +80\, ^{\circ}$ away from the antenna center (see red dashed line in figure~\ref{torpex_sketch} (a)).
The density is obtained from measurements of the ion saturation current $I_{i,sat}$, which is collected with a probe tip polarized at $V_{bias} = -60$~V. We then use $n = |I_{i,sat}|/ (\alpha_0 e A \sqrt{e T_e/m_i} \beta$), with $\alpha_0=0.5$ a coefficient taking into account the effect of the pre-sheath~\citep{book_Lieberman, Furno_2014}, $e$ the electron charge, $A$ the probe tip surface, $T_e$ the electron temperature assumed here constant at 4~eV (measurements in similar conditions, not shown, yield $T_e \sim 2-5$~eV), $m_i$ the ion mass, and $\beta = (1 - \gamma (V_{bias} - V_f))$ a coefficient accounting for the sheath expansion. We use here $\gamma \sim 0.05$, based on results from previous studies in TORPEX~\citep{Theiler_2011}.

Measurements of the turbulent transport are performed with a 3-tip Langmuir probe. A central tip measures the ion saturation current in time, from which the density $n(t)$ is evaluated assuming $T_e\approx 4$~eV. The floating potential $V_f(t)$ is simultaneously measured using two other tips placed on each side and 4~mm away from the central tip along $Z$. This provides an evaluation of $E_{\theta}(t)$.
The turbulent transport is then computed as $\Gamma_{turb} = \langle \tilde{n} \tilde{v}_{E \times B} \rangle$, with $v_{E\times B} = \frac{\vec{E}\times \vec{B}}{B^2} \cdot \hat{X} = E_{\theta} /B_{\phi}$, $\tilde{n} = n - \langle n \rangle$, $\tilde{v}_{E \times B} = v_{E \times B} - \langle v_{E \times B} \rangle$, and $\langle\cdot\rangle$ denoting the time average over the measurement's duration.

\begin{figure}
    \centering
    \includegraphics[width = 0.7\columnwidth, trim={0in 0in 0in 0in},clip]{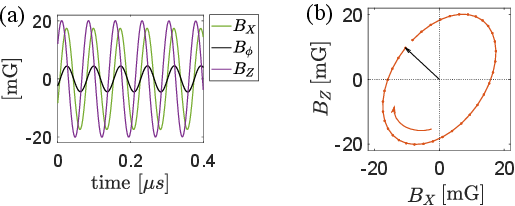}
    \caption{a) Typical example of the magnetic fluctuations measured with a B-dot over a few 13.56 MHz periods. b) Example of the local polarization and eccentricity obtained with the poloidal components of the helicon wave.}
    \label{fig::Bdot_example}
\end{figure}

A magnetic (dubbed B-dot) probe is used to measure the magnetic field fluctuations along all axes $(\hat{X}, \hat{\phi}, \hat{Z})$ at any selected location. 
The B-dot probe consists of three orthogonal 5~mm $\times$ 5~mm square coils, each made of 10 loops of 0.2 mm coated copper wire. B-dot measurements of the magnetic field fluctuations $\vec{B}(t) = (B_X(t), B_{\phi}(t), B_Z(t))$ are performed over 1 ms, at a sampling frequency of 500 MHz.
An example of B-dot measurement is presented in figure~\ref{fig::Bdot_example}.  The local polarization and eccentricity of the poloidal component of the wave is obtained by plotting $B_Z$ as a function of $B_{X}$ during one period of the 13.56 MHz fluctuations (corresponding to 74~ns, i.e. 37 time steps in our measurements), as shown in figure~\ref{fig::Bdot_example} (b).
A reference voltage signal is measured near the antenna input. This provides a phase reference that is used to reconstruct time-varying 1D or 2D profiles of the magnetic fluctuations from B-dot measurements that are performed successively at various positions in any given set of conditions. Details on the B-dot probe design, calibration, and data treatment are provided in~\citep{Vincent_2024}.

B-dot measurements were performed along $X$ and at $Z=0$~cm, at the toroidal angles $\phi \simeq -80 \, ^{\circ}$ and $\phi \simeq +100 \, ^{\circ}$ away from the antenna center, as shown with purple dashed lines in figure~\ref{torpex_sketch}.  In the present study, where many control parameters are explored, most measurements consist of 1D scans along $X$ that are relatively fast to perform. The B-dot probe was nonetheless mounted on a 2D movable system, allowing measurements of the magnetic field fluctuations across a poloidal section of TORPEX.  An example of a 2D scan will be presented in section~\ref{sec::helicon_identification}, to support the helicon mode identification and shed light on the physical interpretation of the radial 1D scans.
Finally, the amplitude of the magnetic fluctuations measured at a given location will be interpreted as the wave amplitude at this location. Note that this would not necessarily be the case for a standing wave, for which a measured amplitude would depend not only on the wave amplitude but also on the toroidal wavelength and the toroidal location of the probe. However, a set of two B-dot probes, placed at different toroidal locations, was used to perform simultaneous measurements of the helicon wave in all the parameters that are explored in this article. None of these simultaneous measurements were in phase or in opposition of phase, excluding the presence of standing waves in this work.
Note that these simultaneous B-dot measurements, not shown here, will be the subject of a future dedicated study on toroidal helicon wavelengths. 

\section{Helicon mode identification}
\label{sec::helicon_identification}

We now present an example of a 2D B-dot measurement, compare it to a simple model, and unambiguously identify a $m=+1$ helicon mode. We show that this identification can be done with a 1D measurement, which will then serve as a reference for the remainder of this article.

For clarity, we emphasize that the term "helicon" is usually used in the context of low-temperature plasma devices to refer to the specific shape that whistler waves adopt in a bounded environment~\citep{Boswell_1970}, and only more recently in the fusion community to the fast wave branch of the whistler waves~\citep{Pinsker_2015}.
As a consequence, the results presented in this section in terms of polarization, poloidal shape, and identification of the helicon mode, cannot be directly compared to the studies of whistler waves in a tokamak.

\subsection{Helicon mode characteristics across a poloidal section}

In the following, we present a simplified model that will help in the interpretation of the experimental data.
The model is derived in~\citep{Chen_1991} and comprises the simplifying hypotheses of a cylindrical geometry, uniform plasma density, and no collisions between electrons and neutrals. With a few additional assumptions (see appendix~\ref{appendix::simple_modeling}) and using a cylindrical coordinate system ($x$, $y$, $z$), the Maxwell equations and momentum conservation of the electrons yield:
\begin{align}
    \bm{\nabla} \times \bm{B} = \alpha \bm{B},
\end{align}
\label{eq::B=alphaB}
where $\alpha = \frac{\omega}{k_z} \frac{\mu_0 e n}{B_0}$, with $\omega$ and $k_z$ the helicon frequency and axial wave number respectively, $\mu_0$ the vacuum permeability and $B_0$ the external magnetic field. Applying perfectly conducting boundary conditions yields an analytic solution for the spatial profile of the helicon mode amplitude, which can be found in ~\citep{Chen_1991, book_Chabert}, and that is reproduced in appendix~\ref{appendix::simple_modeling}.
Note that the toroidal wavelength that is set in the model is $\lambda_z = 1.5$~m. Measurements in TORPEX of the helicon toroidal wavelength $\lambda_{\phi}$ (not shown here) indeed show that we approximately have $\lambda_{\phi} \in [1, 2]$~m for the range of plasma parameters used in the present work.
As discussed in earlier work~\citep{Yasaka_1994, Sakawa_1996, Sudit_1996}, the dominant mode in cylindrical helicon discharges is $m=+1$. 
This mode is therefore of particular importance, and we focus on it in our model.

\begin{figure}
    \centering
    \includegraphics[width = 0.95\columnwidth, trim={0in 0in 0in 0in},clip]{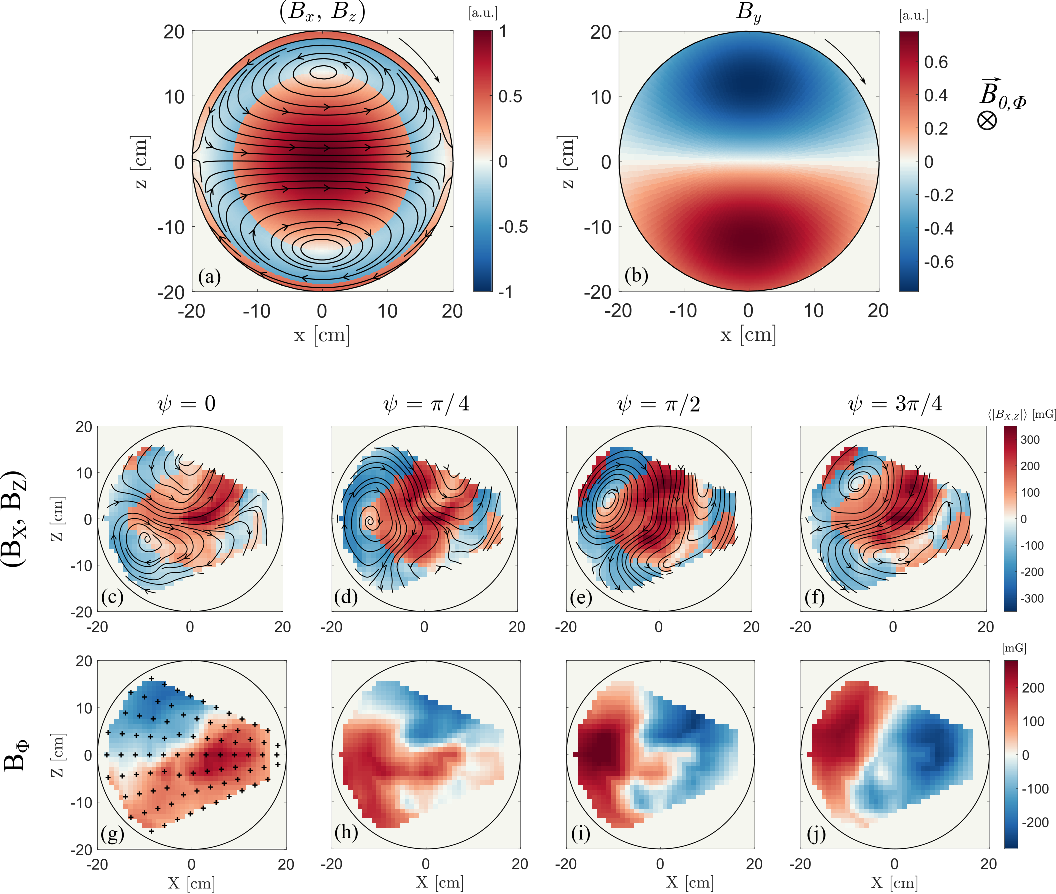}
    \caption{(a-b) Modeling of the helicon mode $m=+1$ in the case of a cylindrical geometry and uniform density. Amplitude of the magnetic helicon wave components perpendicular (a) and parallel (b) to the external magnetic field at the phase $\Psi = 0$. The sign of the colorbar in (a) indicates the sign of the helicon wave polarization as carried by the poloidal component (red is positive with respect to $B_0$, blue is negative), while the colorbar sign in (b) is the sign of $B_y$. (c-j) Experimental magnetic fluctuations measured for an argon base pressure $p_0 \sim 10^{-4}$~mbar, an antenna power of $P_{H}=1$~kW and $B_0 = 300$~G in the SMT configuration, at the toroidal location $\phi \simeq + 100 \, ^{\circ}$, and at four phases of the RF period $\Psi \in [0; \frac{\pi}{4}; \pi ; \frac{3\pi}{4}]$. (c-f) Experimental poloidal component ($B_X$, $B_Z$) amplitude, streamlines (black lines), and polarization (red is positive, blue is negative). (g-j) Experimental toroidal component $B_{\phi}$.}
    \label{fig::streamlines_2D_scan}
\end{figure}

The amplitude of the poloidal and parallel (i.e. along $\hat{\phi}$) magnetic field components of the model helicon mode $m=+1$ are plotted in figure~\ref{fig::streamlines_2D_scan} (a) and (b) respectively. 
In this case, the cylindrical coordinate system ($x$, $y$, $z$) is oriented such that the external magnetic field $B_0$ is along $\hat{y}$. This facilitates the comparison with the toroidal coordinate system ($X$, $\phi$, $Z$) that is used in TORPEX, where $B_0$ is along $\hat{\phi}$. 
An important feature of the $m=+1$ mode is that its local polarization changes sign from the center, where it is positive, to the edge, where it becomes negative, as is shown in figure~\ref{fig::streamlines_2D_scan} (a).
At the edge, within a thin radial layer, the polarization becomes positive again.
On the other hand, the axial component $B_y$ of the mode $m=+1$ (figure~\ref{fig::streamlines_2D_scan} (b)) has a simple structure, being in the direction of $B_0$ in one half of the poloidal section, and opposite to $B_0$ in the other half.
For comparison and completeness, the magnetic fields of the helicon modes $m \in \{-1; 0; 2\}$ from this simple model are provided in appendix~\ref{appendix::simple_modeling}.

An example of magnetic fluctuations measured in TORPEX across the poloidal section at $\phi \simeq + 100 \, ^{\circ}$ is presented in figure~\ref{fig::streamlines_2D_scan} (c-j), for four phases of the RF period $\Psi \in [0; \frac{\pi}{2}; \pi ; \frac{3\pi}{4}]$. These phases correspond to a quarter of a period and are well representative of the evolution of the 2D wave pattern over time.
This measurement was done for an antenna power of $P_{H}=1$~kW, without the magnetron, and a magnetic field of 300~G in the SMT configuration. These conditions are chosen for the stability of the plasma parameters, whose dynamics does not affect the helicon wave, which is best suited for reference comparison with modeling. As will be seen in section~\ref{fig::mag600_bc200_n_Vf}, this is not the case in all experimental conditions, and plasma density fluctuations can strongly impact the helicon wave pattern.
The measurement positions are shown in black crosses by figure~\ref{fig::streamlines_2D_scan} (g), and the resulting data is interpolated on a grid of size 1 cm $\times$ 1 cm for better visualization. Although slightly distorted compared to the modeling shown in figure~\ref{fig::streamlines_2D_scan} (a,b), all the elements that can be extracted from this 2D scan (amplitude of the components of $B$, polarization, pattern of streamlines) unambigously indicate an $m=+1$ helicon mode.
The observed distortion could be an effect of the non-uniformity of the plasma density or the toroidicity, which are the main missing elements of the simplified modeling that is looked at here. This will be investigated in future work.

\subsection{Mode identification from 1D scans}

\begin{figure}
    \centering
    \includegraphics[width = 0.95\columnwidth, trim={0in 0in 0in 0in},clip]{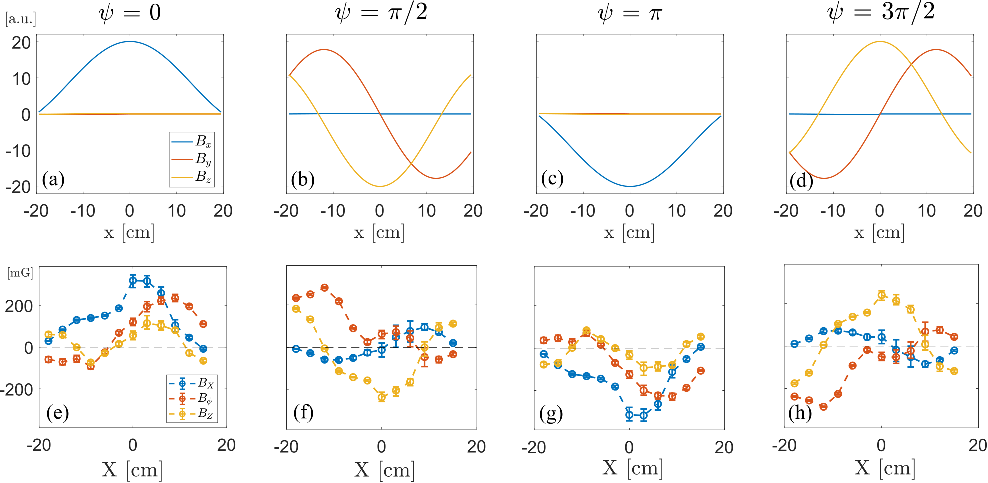}
    \caption{(a-d) Modeling of the amplitude for the magnetic components of the helicon wave across $x$ and at $z=0$, for phases of the waves $\Psi \in [ 0; \frac{\pi}{2}; \pi ; \frac{3\pi}{2} ]$. (e-h) Magnetic fluctuations measured in TORPEX for an argon base pressure $p_0 \sim 10^{-4}$~mbar, an antenna power of $P_{H}=1$~kW and $B_0 = 300$~G in the SMT configuration, at the toroidal location $\phi \simeq + 100 \, ^{\circ}$, and averaged at four phases of the RF period $\Psi \in [0; \frac{\pi}{2}; \pi ; \frac{3\pi}{2}]$.}
    \label{fig::modeling_mode_1}
\end{figure}

Since most of our measurements in TORPEX consist of 1D scans across the plasma column, as it is also commonly done in many experimental devices, we show in figure~\ref{fig::modeling_mode_1} (a-d) the model profiles of ($B_x, B_y, B_z$) across $x$ at $z=0$, and at various phases $\Psi \in [0; \frac{\pi}{4}; \pi ; \frac{3\pi}{2}]$ of the wave evolution along half a period.
The change in local polarization observed in figure~\ref{fig::streamlines_2D_scan} (a), since it is due to the poloidal component relative signs, can also be seen in figure~\ref{fig::modeling_mode_1} (a-d). At a given phase, the sign of $B_x$ remains radially homogeneous (blue curves), whereas $B_z$ changes sign at $x= \pm14.5$~cm (yellow curves).
The structure of $B_y$ of a mode $m=+1$ is also easily identified with a one-dimensional profile along $x$ (see figure~\ref{fig::modeling_mode_1} (a-d), red curves). 
The measured profiles of $(B_X, B_{\phi}, B_Z)$ are also plotted for four phases of the RF period $\Psi \in [0; \frac{\pi}{4}; \pi ; \frac{3\pi}{2}]$, in figure~\ref{fig::modeling_mode_1} (e-h). These profiles are obtained by averaging the instantaneous profiles at these phases over the entire duration of the 1~ms measurements (corresponding to over 10'000 RF cycles), with the error bars representing the standard deviation of the instantaneous profiles from their averaged values.
Note that these error bars are less than a few percent for most of the points, reflecting the great stability of the helicon wave. 
The amplitude profile and time evolution of the magnetic fluctuation components match well with the simplified model of a mode $m=+1$. While $B_X$ alternates between being positive at $\Psi=0$ and negative at $\Psi=\pi/2$, $B_Z$ is negative in the center and positive at the edge at $\Psi=\pi/2$, and the inverse at $\Psi=3\pi/2$. The evolution of $B_Y$ also follows the pattern of mode $m=+1$ for $\Psi=\pi/2$ and $\Psi=3\pi/2$. However, when a magnetic field component is flat and equal to zero in the model, the corresponding measured fluctuations have non-zero values. For the components $B_X$ and $B_Z$ these are small, compared to the component amplitudes, but for $B_Y$ the measured fluctuations are of the same order for all the phases displayed. Note that the evolution of components $B_X$ at $\Psi = \{\pi/2; 3\pi/2\}$ and $B_Z$ at $\Psi = \{0; \pi\}$ could point to a mode $m=0$, but the measurement of $B_Y$ is very different from the strong and centered $B_Y$ oscillation of a mode $m=0$ in the simple model. This could be due to the model not being representative of the experiment, in terms of the predicted theoretical mode shapes.
Actually, these discrepancies with the simple model may be understood from looking back at the 2D scan in figure~\ref{fig::streamlines_2D_scan}. The distortion of the mode makes it impossible to measure a flat zero profile across $Z=0$.
This is for example easily seen with $B_Z$ and $B_Y$ at $\Psi=0$, which are theoretically zero at $Z=0$ (see figure~\ref{fig::streamlines_2D_scan}). These components are clearly non-zero in the measurement, as seen in figure~\ref{fig::instantaneous_profile_magnetron_H2} (c) and (g) respectively, because of only approximate alignment of the 2D structure of the mode with the horizontal.

The 1D profiles display the structure and time evolution of $B_X$, $B_Y$ and $B_Z$ that are characteristic of the simple model of mode $m=+1$, and could not be reproduced with any other mode, nor by including a linear combination of other modes.
Yet, while the 2D scan clearly indicates a $m=+1$ mode, some features of the 1D profiles don't match the $m=+1$ mode characteristics, and could even misleadingly indicate the presence of another mode.
This shows that helicon mode identification based on 1D profiles of individual components of $\vec{B}$ might not be, in general, a robust identification method.

\begin{figure}
    \centering
    \includegraphics[width = 0.95\columnwidth, trim={0in 0in 0in 0in},clip]{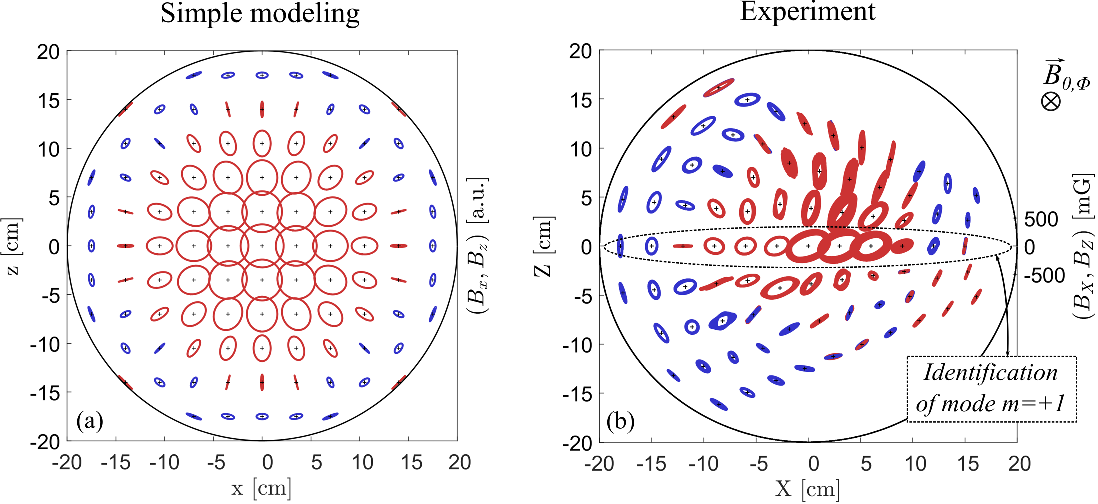}
    \caption{Local polarization and eccentricity of a helicon mode obtained from a full period of the poloidal components' evolution (see figure~\ref{fig::Bdot_example} b). Comparison between (a) a simple model of the helicon mode $m=+1$ assuming uniform density and cylindrical geometry and (b) an experimental measurement at $B=300$~G, $P_{H}=1$~kW, and in the SMT configuration in TORPEX. In (b), at each location, full periods of the evolution of ($B_X$, $B_Z$) at 10 different times across the 1~ms duration of the measurement are plotted. Red corresponds to a positive polarization with respect to $\hat{y}$ (a) or $\hat{\phi}$ (b), and blue to a negative one.}
    \label{fig::ellipses_2D}
\end{figure}

Another way to represent the spatio-temporal evolution of ($B_X, B_Z$) is to plot at each measurement location the evolution of $B_Z(t)$ as a function of $B_X(t)$ over a wave period, as it is done in figure~\ref{fig::Bdot_example} (b). 
This way of plotting shows in a single image the polarization, eccentricity and amplitude of the poloidal component of the magnetic fluctuations that are needed for mode identification, without the need for a full 2D scan. It can be used to distinguish between positive and negative $m$ modes using the polarization pattern, and the mode number using the profile amplitude. Figure~\ref{fig::ellipses_2D} (a, b) shows this evolution of ($B_X(t), B_Z(t)$) across a poloidal plane, in the case of the mode $m=+1$, for the modeling and the measurement, respectively.
Looking at the simplified model (figure~\ref{fig::ellipses_2D} (a)), in addition to the local change in polarization's sign, we note a shape that is circular inside the area of positive polarization, elliptical at the edge where the polarization is negative, and almost linear at the interface between both areas.
These elements are also clearly visible on the experimental magnetic fluctuations that are used in this section as a reference (figure~\ref{fig::ellipses_2D} (b)).
This compact and convenient way of representing ($B_Z$, $B_X$) will therefore be used in the rest of this article to look at profiles across $X$, providing a clear identification of a dominant helicon mode $m=+1$ across all our datasets, as will be seen in sections~\ref{sec::magnetron_generated_plasma}, \ref{sec::power_scan}, \ref{sec::mag_field_scans}.

\section{Helicon waves launched in a background magnetron-generated plasma}
\label{sec::magnetron_generated_plasma}

As a first step for the characterization of toroidal helicon waves and propagation properties, the magnetron is used to produce a background plasma in TORPEX, on top of which helicon waves are launched by the helicon antenna at low power. First measurements of this type were performed in hydrogen and in the SMT configuration, as reported in \citep{Vincent_2024}. In this section, we extend these measurements to both PT and SMT configurations, and to argon plasmas. 

The aim of using the helicon antenna at low power on a background magnetron-generated plasma is to limit the impact of the helicon source on the plasma equilibrium. The direct effect of the toroidal plasma configuration on the helicon waves can then be studied, with a limited reciprocal impact of the helicon waves on the plasma equilibrium parameters.

\subsection{Impact of helicon antenna on the plasma parameter mean profiles}
\label{subsec::magnetron_impact_helicon_plasma}

\begin{figure}
    \centering
    \includegraphics[width = 0.8\columnwidth, trim={0in 0in 0in 0in},clip]{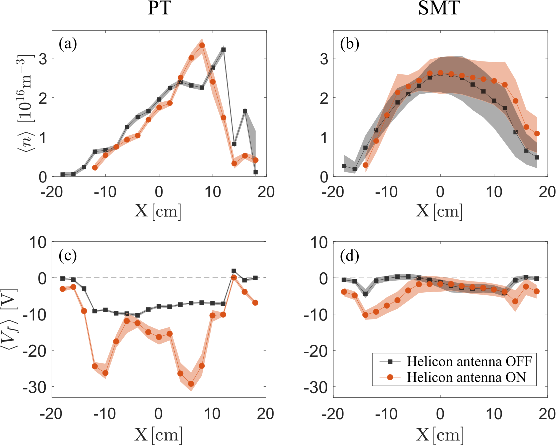}
    \caption{Density and floating potential radial profiles measured by a Langmuir probe at the toroidal location $\phi \sim + 80 \, ^{\circ}$, with $P_{MAG} = 600$~W, with and without the helicon antenna at $P_{H} = 200$~W. Measurements were performed in the magnetic configurations PT (a,c) and SMT (b,d).}
    \label{fig::mag600_bc200_n_Vf}
\end{figure}

A background plasma is generated with the magnetron at a power of 600 W in argon and at a base pressure of $p_0 \sim 1 \times 10^{-4}$~mbar. We set $B_0 \approx 780$~G, and for the SMT configuration a vertical field of $B_{ver} \approx 20$~G is added. 
The density and floating potential are measured with a Langmuir probe across $X$ at the vertical position $Z = 0$, and average values are presented in figure~\ref{fig::mag600_bc200_n_Vf}. The shaded areas represent the standard deviation of the fluctuations of $n$ and $V_f$ relative to their mean values.
We note that at this pressure and power, the plasma parameters are conserved along the toroidal direction. This has been verified by simultaneous measurements of two 2D arrays of Langmuir probes placed at opposite toroidal locations in TORPEX with a magnetron-generated plasma (not shown here). We can therefore consider that these profiles measured at $\phi \sim + 80 \, ^{\circ}$ with the magnetron only (black curves of figure~\ref{fig::mag600_bc200_n_Vf}) are well representative of the plasma parameters along the toroidal direction.

In the PT configuration (figure~\ref{fig::mag600_bc200_n_Vf} (a,c)), the plasma density is peaked on the low-field side (LFS), reaching $n\approx 3 \times 10^{16}$~m$^{-3}$ around $r\approx 12$~cm. This is consistent with the presence of strong outward drift due to charge separation.
The floating potential is constant at $\approx - 10$~V in the region inside the limits defined by the antenna glass tube, i.e. for $r \leq 15$~cm, and gets close to zero at the edges.
With the addition of a vertical magnetic field in the SMT configuration (figure~\ref{fig::mag600_bc200_n_Vf} (b,d)), the plasma confinement is improved, and $n \approx 2.5 \times 10^{16}$~m$^{-3}$ around the center, with a more homogeneous spreading of the density in the range $r\in[-10 ; 10]$~cm. In SMT, $V_f$ is close to zero across the entire profile.

Helicon waves are then launched with the helicon antenna operating at a power of $P_{H} = 200$~W, a third of the magnetron power. Compared to the situation with the antenna off, in the PT configuration the density peak is slightly moved to $r\approx 8$~cm, and in the SMT configuration the relatively high density level of $\approx 2.5 \cdot 10^{16}$~m$^{-3}$ extends to a slightly wider region in $r\in[-10 ; 10]$~cm, as shown in figure~\ref{fig::mag600_bc200_n_Vf} (a, b). Overall, the density profiles are little affected. This is interesting because it shows a regime where helicon waves are excited with a weak contribution to the plasma ionization, which enables the study of the properties of helicon waves independently of their complex energy deposition mechanisms~\citep{Chen_1991, Canases_2016, Guittienne_2025}.
The floating potential, on the other hand, is clearly modified when the antenna is turned on. The value of $V_f$ is reduced around $r \approx 10$~cm, close to the antenna glass tube wall, by $ \approx 20$~V in the PT configuration and by $ \approx 5\text{-}10$~V in the SMT one. This may be attributed to an increase in $T_e$ from electron heating by the helicon antenna close the glass tube.

\subsection{Dominant helicon mode propagation}
\label{subsec::magnetron_propagation_helicon}

\begin{figure}
    \centering
    \includegraphics[width = 0.95\columnwidth, trim={0in 0in 0in 0in},clip]{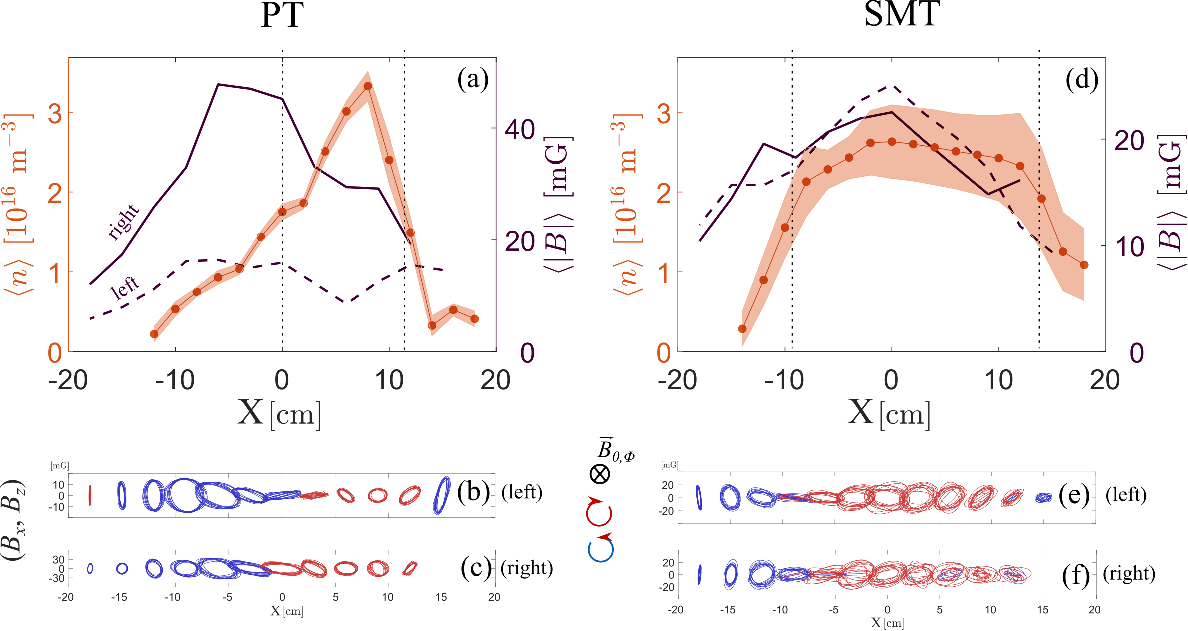}
    \caption{(a,d) Comparison of the plasma density to the helicon wave amplitude measured at the toroidal angles $\phi \simeq -80 \, ^{\circ}$, i.e. "left" of the antenna (plain purple curves), and $\phi \simeq +100 \, ^{\circ}$, i.e. "right" of the antenna (dashed purple curves), all measured along $X$, at $Z=0$, for $P_{MAG} = 600$~W and $P_{H} = 200$~W. Poloidal component of helicon waves measured with a B-dot probe along $X$, at $Z=0$, on the left (b,e) and right (e,f) of the helicon antenna, for magnetic configurations of PT (b,c) and SMT (e,f). Red corresponds to a positive polarization of the wave with respect to the toroidal axis $\hat{\phi}$, blue to a negative one.}
    \label{fig::comparison_helicon_density_Ar}
\end{figure}

We now focus on the helicon waves launched by the antenna. 
The magnetic fluctuations are measured across $X$ at $Z=0$, on both sides of the antenna at $\phi \simeq -80 \, ^{\circ}$ and $\phi \simeq + 100 \, ^{\circ}$. Figure~\ref{fig::comparison_helicon_density_Ar} (b,c) and (e,f) show the local polarization and eccentricity of the poloidal component of $B$ in PT and SMT configurations respectively, superimposing full periods of the evolution of ($B_X$, $B_Z$) at 10 different times across the 1~ms measurement. 
For comparison, the plasma density profiles are also plotted in figure~\ref{fig::comparison_helicon_density_Ar} (a,d).

The local polarization changes sign along $X$, and follows the plasma density, with a central positive polarization in the plasma bulk and a negative polarization at the plasma edges. For better visualization, vertical dashed lines are plotted on top of the density profile, at the locations of changes in local polarization. 
We also see that the eccentricity of ($B_X, B_Z$) is close to 0 (i.e. close to a circular polarization) at the center of these areas of positive and negative local polarization, whereas it is close to 1 (i.e. close to a linear polarization) at the transition between them. 
With these elements, a helicon mode $m=+1$ can be clearly identified in both magnetic configurations and at each side of the antenna.
The instantaneous profiles for each component of $B$, at four phases of the time evolution of the wave, are provided in appendix~\ref{appendix::instant_profiles} and support this identification.

The helicon mode mean amplitudes $\langle |B| \rangle  = \big \langle \sqrt{ B_X^2(t) + B_{\phi}^2(t) + B_Z^2(t) } \big \rangle_t$ are plotted in figure~\ref{fig::comparison_helicon_density_Ar} (a,d). The profiles of $\langle |B| \rangle$ are centered in both magnetic configurations and seem uncorrelated with the density profile or with the polarization pattern of the mode.
In the PT configuration (figure~\ref{fig::comparison_helicon_density_Ar} (a)), the helicon mode amplitude is three times larger on the right of the antenna (purple plain curve) than on the left (purple dashed curve). This indicates a preferential propagation of the wave in the direction of the confining magnetic field $\vec{B}_{0} \propto \hat{\phi}$, and a strong damping of the wave after propagation along the entire torus of TORPEX.
With the addition of a small vertical field in the SMT configuration (figure~\ref{fig::comparison_helicon_density_Ar} (b)), the amplitude of the helicon wave on the right and on the left of the antenna are almost identical. This either shows that a wave is launched on one side of the antenna and propagates with very little damping along the torus of TORPEX until it reaches the other side, or that two $m=+1$ waves are launched, one on each side of the antenna.
With the symmetric design of the helicon antenna mounted on TORPEX, the latter is expected. Notice that in this case the mode $m=+1$ launched along $B_0$ would have a right-handed spatial helicity, whereas the one launched in the opposite direction would have a left-handed helicity. Distinguishing between those two cases would require a series of B-dot measurements along the axial direction $\hat{\phi}$, providing a spatial reconstruction of the helicon mode as was done in~\citep{Guittienne_2021}, for which TORPEX is not equipped at the moment.

All these measurements were also performed in hydrogen (part of which are reported in~\citep{Vincent_2024}, and in the appendix~\ref{appendix::instant_profiles}). The results obtained in hydrogen are similar to those obtained in argon that are reported in this section. This shows that the toroidal helicon wave behavior, in terms of propagation, mode number, and polarization shape, is not dependent on the ion mass. This could be expected by the fact that helicon wave behavior is governed by the electron dynamics, and is consistent with TORPEX experimental conditions where the ions can be considered cold and where $\omega_{ci} \ll \omega$ in both hydrogen and argon, with $\omega$ the excited helicon wave at 13.56 MHz.

We finally point out that during the entire measurement lasting one millisecond, the helicon wave is very stable in the PT configuration (figure~\ref{fig::comparison_helicon_density_Ar} (b,c)), whereas strong amplitude and shape variations are observed in the SMT case (figure~\ref{fig::comparison_helicon_density_Ar} (e-f)).
We argue that this is due to the low-frequency waves that develop in TORPEX and interact with the helicon waves, as investigated next.

\subsection{Interaction with low-frequency waves}
\label{subsec::magnetron_interaction_LF_waves}

Helicon wave nonlinear interactions have been reported in the literature, for instance in the case of a parametric decay involving ion acoustic waves~\citep{Kline_2003}, or of helicon second harmonic generation~\citep{Karimov_2026}.
However, the interaction of helicon waves with low-frequency drift or interchange waves, which have a critical contribution to turbulence in the edge of fusion plasmas, has never been reported to our knowledge.
In TORPEX, the SMT magnetic configuration is prone to the development of low-frequency (LF) waves of the interchange or resistive-drift type, which have been the focus of previous studies~\citep{Poli_2008, Ricci_2010}.

\begin{figure}
    \centering
    \includegraphics[width = 0.95\columnwidth, trim={0in 0in 0in 0in},clip]{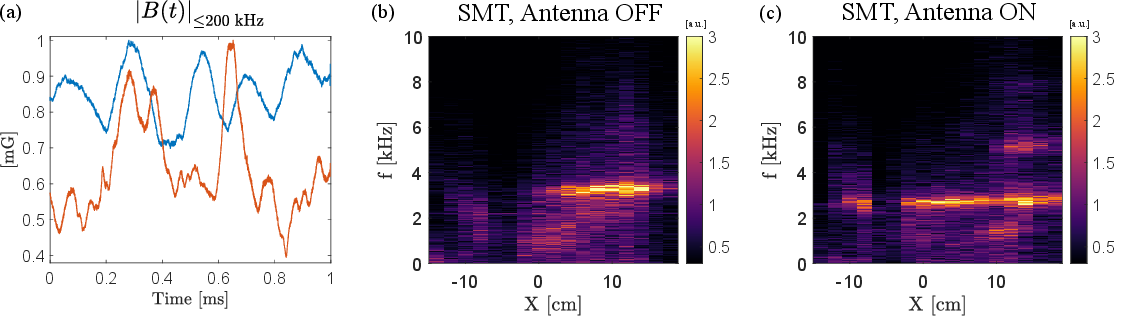}
    \caption{(a) Envelope of the magnetic fluctuations measured at $X=9$~cm in PT and SMT configurations. (b,c) Spectra of the density fluctuations as a function of the position $X \in [-14:2:18]$~cm. Measurements were performed at ($Z=0$, $\phi \simeq + 80 \, ^{\circ}$) in argon plasma with $B_0=780$~G, for a magnetron power of $600$~W, and with the helicon antenna power (b) turned off and (a,c) turned on at $P_{H} = 200$~W.}
    \label{fig::NL_interaction_helicon_LF_BC_on_off}
\end{figure}

Figure~\ref{fig::NL_interaction_helicon_LF_BC_on_off} (a) shows the envelope of $|B(t)|$ in a 600~W magnetron-generated plasma with antenna power of 200~W, in PT and SMT configurations. This envelope is obtained by applying a low-pass filter to the B-dot signals, with a cut-off frequency of 200~kHz. In the PT configuration, where LF waves are weak, the envelope of $|B(t)|$ remains within $\pm 15\%$ of its mean value (blue curve). In contrast, in the SMT configuration where strong LF waves develop, the envelope of $|B(t)|$ is modulated by more than $\pm 50\%$ of its mean value, at a timescale of $\sim$ms corresponding to kHz fluctuations (red curve). Longer B-dot measurements, from which kHz frequencies can be extracted with more accuracy, and simultaneous density measurements (see appendix) reveal that these frequencies indeed correspond to the density fluctuation frequencies.
This provides clear evidence that the LF waves developing in TORPEX have a strong impact on the helicon waves launched in the plasma, with the effect being a modulation of the helicon wave amplitude.

On the other hand, we also observe an impact of the helicon waves on the plasma density fluctuations. 
Figure~\ref{fig::NL_interaction_helicon_LF_BC_on_off} (b,c) shows the spectra of the plasma density fluctuations for each measurement position along the $X$ axis, in the SMT configuration, without helicon waves and for an antenna power of $P_{H}=200$~W.
With the helicon antenna turned off, fluctuations are clearly visible around $4\leq X \leq 14 $~cm at $\approx 3.3$~kHz, which we identify as LF waves.
When the antenna is powered on, these LF waves are still visible at a slightly lower frequency of $\sim 2.6$~kHz, but they are extended to the high-field side (HFS) to $X\approx-8$~cm and into the plasma core to $ -2\leq X \leq 6 $~cm. A harmonic component at $\approx 5.2$~kHz is also generated on the LFS, as well as a lower frequency component at $\approx 1$~kHz.
At the plasma HFS, this enhancement of LF instabilities may be attributed to the slight increase in the plasma density gradient, as can be seen in figure~\ref{fig::mag600_bc200_n_Vf} (b) at $X\sim-10$~cm.
But this is not the case on the LFS, where the density gradient is mostly unchanged, and even slightly reduced at $X\sim10$~cm. The distinct element that can explain the LF wave enhancement from one experiment to the other is the launching of helicon waves. 
At $X=0$, gradients of $n$ and $V_f$ are flat. At this same location, the helicon mode amplitude is at its highest (see figure~\ref{fig::comparison_helicon_density_Ar} (d)), and is therefore identified as the likely source of LF waves at this location, through non-linear energy transfer from 13.56~MHz to the kHz range.

The physical mechanism of this coupling is not entirely elucidated and requires dedicated investigations that are out of the scope of this article.
We note nonetheless that LF waves also appear in the floating potential spectra (not shown here), which indicates a possible energy transfer from the magnetic to density fluctuations through the electric field fluctuations. 
We also note that a similar behaviour of LF wave enhancement by helicon waves is observed with hydrogen plasmas, although not discussed in detail here. Spectra of the density fluctuations in hydrogen plasmas are shown in appendix~\ref{appendix::instant_profiles}.

As the antenna power is increased, further evidence of non-linear coupling between the $13.56$~MHz helicon mode and the $\sim$kHz plasma density fluctuations is observed; this behavior is examined in Section~\ref{sec::power_scan}.

\section{Impact of antenna power on toroidal helicon plasma generation}
\label{sec::power_scan}

We now investigate the ability of the helicon antenna to generate plasma in TORPEX, and the associated helicon waves that are produced. In addition, we explore the link between helicon waves and plasma dynamics, which is usually not investigated in helicon plasma source studies.

\begin{figure}
    \centering
    \includegraphics[width = 0.9\columnwidth, trim={0in 0in 0in 0in},clip]{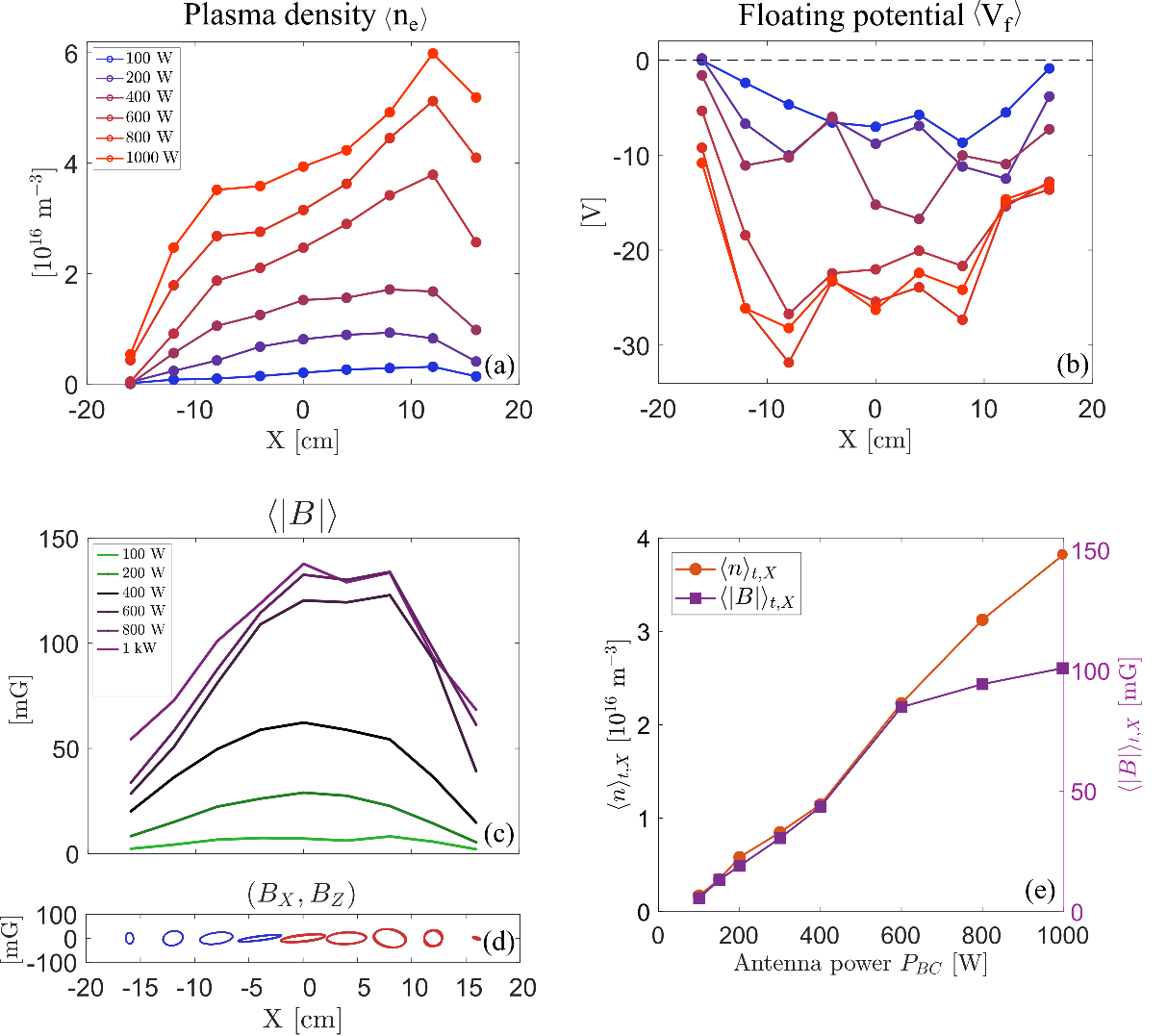}
    \caption{Mean profiles of (a) the plasma density $\langle n \rangle$ and (b) the floating potential $\langle V_f \rangle$ measured at ($Z=0$, $\phi \simeq + 80 \, ^{\circ}$), compared to (c) the mean profiles of the poloidal magnetic fluctuation amplitude $\langle |B_{pol} |\rangle$. (d) Shape of ($B_X, B_Z$) for $P_{H} = 400$~W. (e) Comparison between the integrated profiles of $\langle n \rangle_{t,X}$ and $\langle |B| \rangle_{t,X}$. All measurements are performed in PT configuration at $B_0=500$~G and $p_0 \sim 10^{-3}$~mbar.}
    \label{fig::power_scan_n_Vf_helicon}
\end{figure}

The magnetic configuration is set to PT with $B_0 = 500$~G, and argon at $p_0 \sim 10^{-3}$~mbar.
This pressure is chosen higher than in section~ \ref{sec::magnetron_generated_plasma} because it provides an easier coupling between the antenna and the plasma. In the remainder of this article, the magnetron is not used, and the antenna power is varied within $P_{H} \in [100 ; 1000]$~W.
Profiles of the plasma density and floating potential are measured with the Langmuir probe, along the $X$-axis and at $Z=0$, at the toroidal location $\phi \sim + 80 \, ^{\circ}$, and are shown in figure~\ref{fig::power_scan_n_Vf_helicon} (a,b). The plasma density profiles are asymmetric with a higher density on the low field side (LFS). This is expected with a confining magnetic field that is purely toroidal, causing a vertical charge separation and a global $E\times B$ drift of the plasma towards the LFS. Figure~\ref{fig::power_scan_n_Vf_helicon} (a) also indicates a linear increase of the plasma density with the antenna power. This linear dependence is clearly shown in figure~\ref{fig::power_scan_n_Vf_helicon} (e), where the integrated profile of the plasma density $\langle n\rangle_{t,X} = \frac{1}{2a}\int_X \langle n \rangle_t dX 
$ is plotted as a function of the antenna power (red curve). Notice that with a helicon antenna, abrupt density increases could be expected as the power is raised, as is observed in a number of laboratory helicon devices~\citep{Boswell_1984, Sakawa_1996, Chi_1999, Zhang_2024}. These studies commonly use antennae of diameter $\sim 10$~cm, and report sharp density transitions at antenna powers of $\sim 100-1000$W. In TORPEX, with the antenna of diameter $\sim 30$~cm encompassing a volume approximately ten times larger than the commonly used helicon antennae, we expect sharp density transitions to be possible only at much higher powers of $\sim10$~kW. This will be explored in further studies. Contrary to the density, the floating potential profiles, shown in figure~\ref{fig::power_scan_n_Vf_helicon} (b), display a sharp transition, with values going from $V_f \sim -10$~V at $P_{H} = 400$~W to $V_f \sim -25$~V at $P_{H} = 600$~W. This suggests a sharp increase in electron temperature, likely linked to the helicon wave saturation that is discussed in the following.

The global amplitude of the magnetic fluctuations $\langle |B |\rangle$ is plotted in figure~\ref{fig::power_scan_n_Vf_helicon} (c) as a function of the radius of TORPEX, at $Z=0$ and $\phi \simeq + 80 \, ^{\circ}$, for increasing powers of the helicon antenna. 
The shape of ($B_X, B_z$) is plotted in figure~\ref{fig::power_scan_n_Vf_helicon} (d) for $P_{H} = 400$~W, and is well representative of ($B_X, B_z$) for all the measurements of the power scan (not shown here).
The polarization pattern shows a negative polarization on the HFS and a positive polarization on the LFS, where the density is stronger. In the core of the regions where the polarization is locally positive or negative, it is close to circular. At the boundary between these regions, it is more elongated and close to linear.
These features, also visible in figure~\ref{fig::ellipses_2D}, indicate that the magnetic fluctuations are composed of a dominant $m=+1$ helicon mode, with a polarization that follows the density profiles.
The radial profile of the helicon mode amplitude, however, remains centered around $X=0$ and does not follow the plasma density profile that is asymmetric. This extends the observation that was made in figure~\ref{fig::comparison_helicon_density_Ar} (a), indicating that in a PT magnetic configuration, mean helicon wave amplitude and mean plasma density do not seem correlated.

Another important observation is the saturation of the helicon mode amplitude when the antenna power is increased. The profile of $\langle |B |\rangle$ increases linearly with antenna power up to $P_{H}=600$~W, with a value at the center of $\langle |B|\rangle \simeq 7 $~mG at $P_{H}=100$~W, going up to $\langle |B|\rangle \simeq 120 $~mG at $P_{H}=600$~W. Further raising the power to $P_{H}=1000$~W only increases the helicon amplitude of about $15 \%$, with $\langle |B|\rangle \simeq 140 $~mG at $X=0$~cm. This saturation is highlighted in figure~\ref{fig::power_scan_n_Vf_helicon} (e), where the integrated profile of the helicon amplitude $\langle |B|\rangle_{t,X}$ is plotted as a function of the antenna power (plain purple curve) along with $\langle n \rangle_{t,X}$. The values of $\langle |B|\rangle_{t,X}$ and $\langle n \rangle_{t,X}$ follow each other very closely along a linear trend as a function of the antenna power, up to $P_{H} = 600$~W. Then the helicon amplitude departs from this trend and only increases by $\approx 15\%$, while the antenna power (and plasma density) almost doubles from $600$~W to $1$~kW.

\begin{figure}
    \centering
    \includegraphics[width = 0.9\columnwidth, trim={0in 0in 0in 0in},clip]{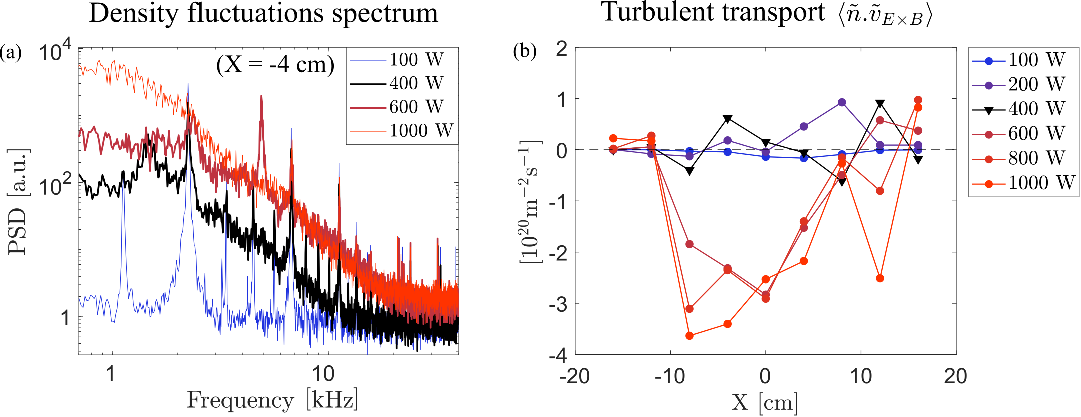}
    \caption{(a) Spectra of the corresponding density fluctuations at $X=-4$~cm. (b) Turbulent transport $\langle \tilde{n}.\tilde{v}_{E\times B} \rangle$ measured at ($Z=0, \phi \simeq + 80\, ^{\circ}$) for helicon antenna powers of $P_{H} \in [100, 1000]$~W.}
    \label{fig::tau_turb_spectra_power_scan}
\end{figure}

This transition is further explored by computing the Fourier spectra of the density fluctuations, as shown in figure~\ref{fig::tau_turb_spectra_power_scan} (a) for the location $X=-4$~cm. The spectra reveal a strong increase in density fluctuations in the kHz range as the power is raised, gaining notably an order of magnitude as $P_{H}$ varies from $400$~W to $600$~W. This is particularly interesting since this position $X=-4$~cm corresponds to a location along $X$ where the plasma density gradient is close to its smallest value, showing that this strong increase in LF fluctuations does not correlate with a density gradient increase. This suggests that the LF fluctuations might be linked to the helicon mode amplitude, through a non-linear energy transfer from $\approx 13.56$~MHz to the kHz range.
For $P_{H}\geq 600$~W, while the mean plasma density keeps increasing, the plasma density fluctuations for frequencies in $[2; 15]$~kHz saturate, as does the helicon wave amplitude. This further tends to indicate that the density fluctuations are powered by the helicons.

The turbulent transport $\Gamma_{turb} = \langle \tilde{n}.\tilde{v}_{E \times B} \rangle$ is also measured for increasing $P_{H}$, across $X$ at ($Z=0$, $\phi \simeq + 80 \, ^{\circ}$), and shown in figure~\ref{fig::tau_turb_spectra_power_scan} (b). 
The measured order of magnitude is $\Gamma_{turb} \sim 10^{19}-10^{20}$~m$^{-2}$s$^{-1}$.
For comparison, an estimate of the ambipolar diffusion coefficient~\footnote{We estimate the ambipolar diffusion coefficient as $D_a \approx (1+\frac{T_e}{T_i})D_i$, with $D_i = v_{th,i}^2/\nu_{ii}$ the ion diffusion coefficient in TORPEX with weakly magnetized ions, where $v_{th,i}=k_B T_i/m_i$ and $\nu_{ii} \simeq \sigma_{ii}^* n_iv_{th,i} \sim10^5$~Hz the cumulative small-angle scattering ion-ion collision frequency~\citep{book_Bellan}.}
in TORPEX gives $D_a \sim 10^2$~m$^2$/s. With $\frac{\partial n}{\partial X} \sim 10^{16} - 10^{17}$~m$^{-3}$ computed from the measurements, the transport due to diffusion can be estimated at $\sim 10^{18} - 10^{19}$~m$^{-2}$s$^{-1}$, i.e. an order of magnitude lower than the turbulent transport, which is therefore dominant across all the measurements.
A sharp and strong transition can nonetheless be observed. At the location $X\in[-10;5]$~cm, the turbulent transport stays within $|\Gamma_{turb}| \lesssim 0.5 \times 10^{20}$~m$^{-2}$s$^{-1}$ for $P_{H}\leq 400$~W. At $P_{H} = 600$~W, it abruptly increases to values of $\Gamma_{turb}\approx -3 \times 10^{20}$~m$^{-2}$s$^{-1}$, and remains in this range for $P_{H} > 600$~W.
The sudden jump of $\Gamma_{turb}$ could be the reason for the saturation of the helicon mode at $P_{H} = 600$~W. On the other hand, the much slower increase of $\Gamma_{turb}$ for $P_{H} \geq 600$~W can be linked to the saturation of the LF fluctuations, which itself appears to result from the helicon mode saturation as discussed earlier.
Without quantitative estimates of non-linear interactions and energy transfers, these strong correlations between helicon mode amplitude, LF fluctuations, and turbulent transport have to be interpreted with care, and would need dedicated studies for more accurate conclusions. 
This set of measurements nonetheless exhibits an overall rich non-linear dynamics between helicon waves and LF instabilities, which has remained so far largely unexplored in the literature to our knowledge, and is experimentally observed for the first time in a toroidal device.

\section{Impact of magnetic field on toroidal helicon plasma generation}
\label{sec::mag_field_scans}

We now focus on the impact of the magnetic field on the helicon mode characteristics and the related impact on the plasma parameters.
We note that the result of the present section also constitutes an exploration of entirely new plasma regimes in TORPEX.
Indeed, in all the previous studies in TORPEX, the use of a magnetron for plasma generation has limited the confining magnetic field to a range of ${B_0\in[650;850]}$~G. With the implementation of a helicon antenna, this constraint is removed. In this section, the value of the confining magnetic field is varied within ${B_0\in[100;  900]}$~G, in both magnetic configurations (PT and SMT). An argon plasma is generated with $P_{H}=1$~kW, at  $p_0 \sim 10^{-3}$~mbar.

\begin{figure}
    \centering
    \includegraphics[width = 0.98\columnwidth, trim={0in 0in 0in 0in},clip]{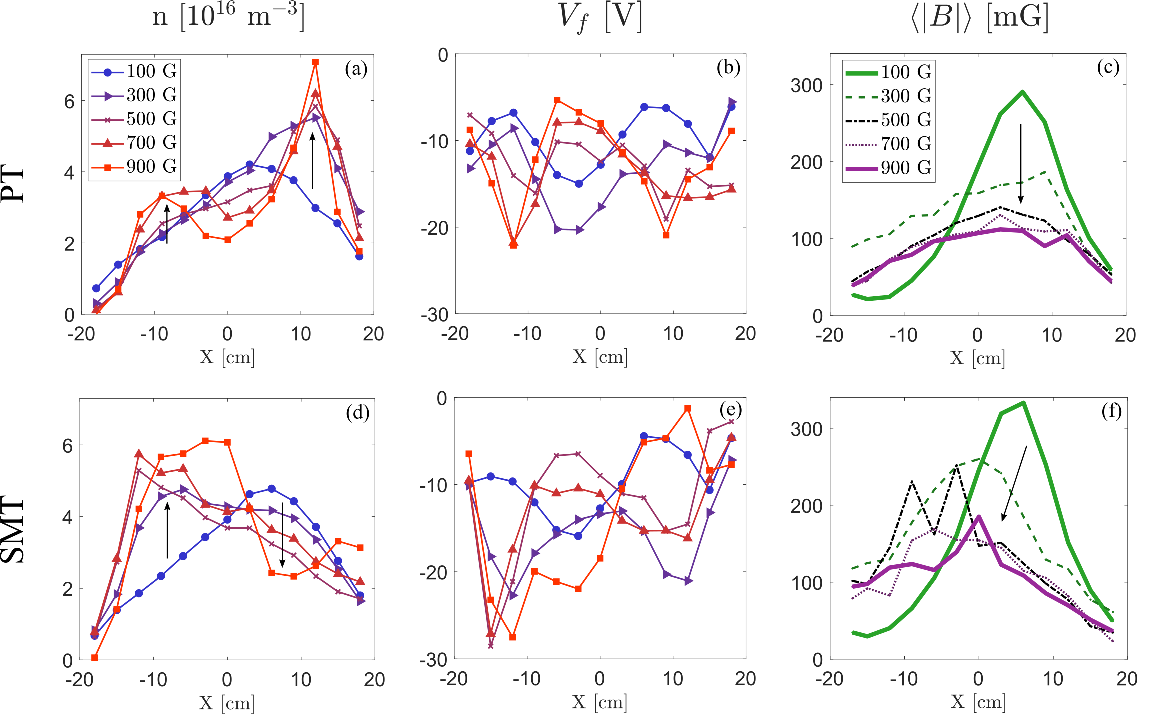}
    \caption{Mean profiles of (a,b) the plasma density $\langle n \rangle$, (d,e) the floating potential $\langle V_f \rangle$ measured at ($Z=0$, $\phi \simeq + 80 \, ^{\circ}$) and (c,f) the magnetic fluctuation amplitude $\langle |B| \rangle$, for a magnetic field of $B_0 \in \{100,300,500,700,900\}$ and with a magnetic configuration of (a,b,c) pure toroidal field and (d,e,f) SMT.}
    \label{fig::mag_scan_n_Vf_B_all}
\end{figure}

The measured mean profiles of the plasma density and floating potential are shown in figure~\ref{fig::mag_scan_n_Vf_B_all} (a-b) and (d-e) for PT and SMT, respectively. In both configurations, increasing the magnetic field has a strong impact on the shape of the plasma density profile, without modifying its mean value along $X$. In PT, the density profile, initially centered around $X=0$~cm for $B_0=100$~G, peaks at $X\simeq \pm 10$~cm as $B_0$ is increased.
This can be understood as an effect of stronger confinement along the field lines, hindering the diffusion of the plasma that is generated close to the antenna at the radius $r=15$~cm.
In SMT, this effect disappears with the addition of the vertical magnetic field. Indeed, in SMT, field lines connect the top to the bottom of the vessel, crossing all radial positions, allowing parallel diffusion to transport plasma radially. The density profile is nonetheless also impacted by the increase of $B_0$, with a density peak gradually moving towards the HFS.

\begin{figure}
    \centering
    \includegraphics[width = 0.95\columnwidth, trim={0in 0in 0in 0in},clip]{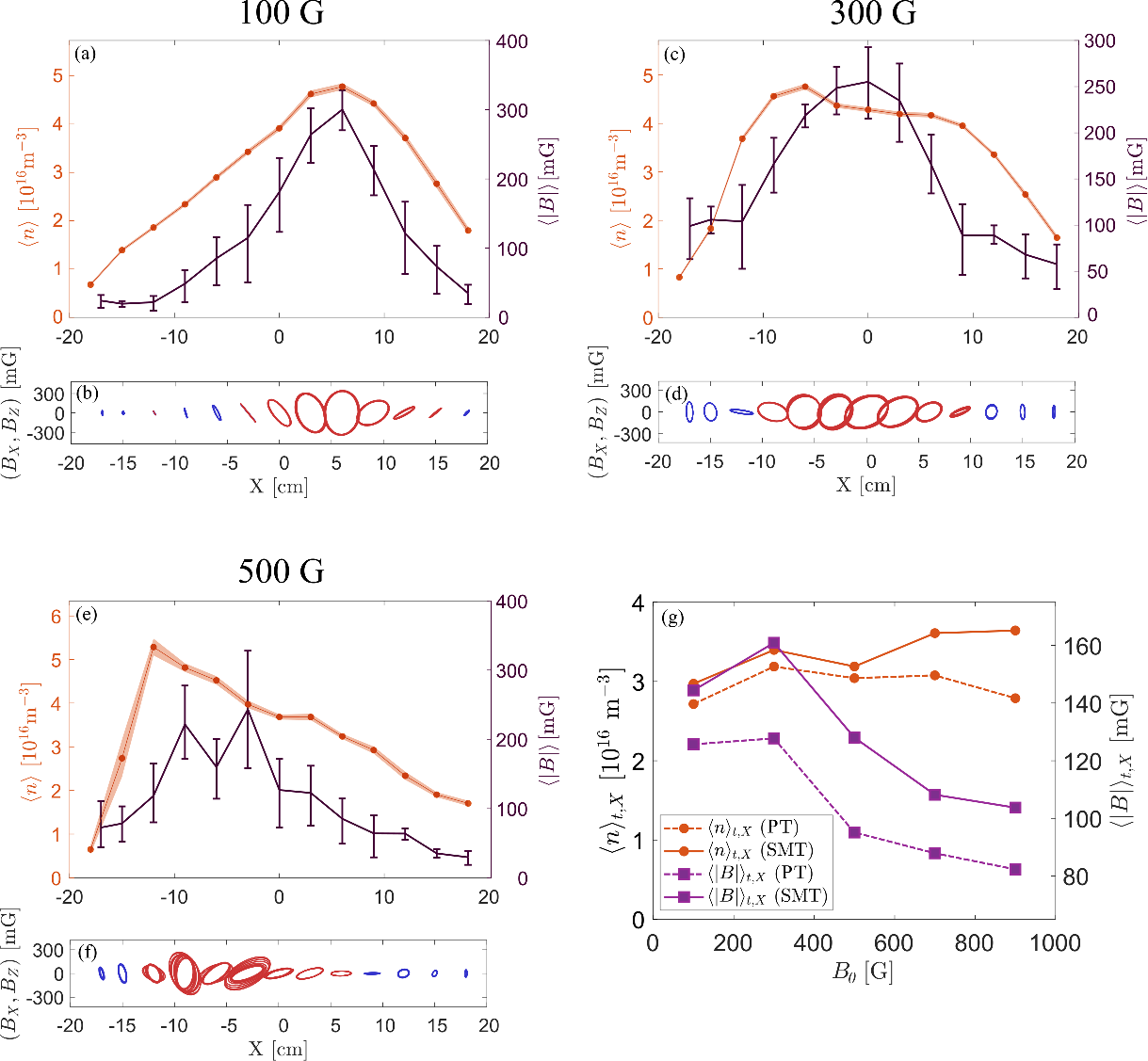}
    \caption{Comparison between (a,c,e) $\langle n \rangle$, $\langle |B| \rangle$ and (b, d, f) the shape of ($B_X, B_Z$) at ten times across the 1~ms measurements, for SMT configuration for (a,b) $B_0 = 100$~G, (c,d) $B_0 = 300$~G and (e,f) $B_0 = 500$~G. (g) Integrated density  $\langle n\rangle_{t,X}$ and magnetic field fluctuations $\langle |B| \rangle_{t,X}$ as a function of $B_0$ in PT and SMT configurations.}
    \label{fig::mag_scan_density_B_SMT}
\end{figure}

In both magnetic configurations, however, the plasma density remains within $\sim2-6 \times10^{16}$~m$^{-3}$ in the bulk of the plasma for $|X| \leq 15$~cm. The fully integrated density $\langle n \rangle_{t,X}$ plotted in figure~\ref{fig::mag_scan_density_B_SMT} (g) exhibits an almost constant value around $\sim 3 \times 10^{16}$~m$^{-3}$ from $B_0 = 100$~G to $B_0 = 900$~G.
The floating potential, like the density, remains within a constant range of values as $B_0$ is increased, at $\langle V_f \rangle \sim -15$~V, while the profile shape becomes more peaked on the edges. This shows that $B_0$ has a strong impact on the plasma equilibrium, as could be expected, but does not seem to change the overall plasma generation efficiency of the helicon antenna.

Measurements of the magnetic fluctuations are also performed for $B_0 \in [100; 900]$~G in the PT and SMT configurations, and profiles of $\langle |B| \rangle$ are shown in figure~\ref{fig::mag_scan_n_Vf_B_all} (c,f).
In both cases, the profile of $\langle |B| \rangle$ peaks around $X\sim5$~cm for a low confining magnetic field of $B_0=100$~G. When $B_0$ is increased, $\langle |B| \rangle$ is lowered, and at $B_0=900$~G we have $\langle |B| \rangle \sim 100$~mG with approximately on-axis centered profiles.
Note that in PT, $\langle n \rangle$ and $\langle |B| \rangle$ are not correlated, whereas they are in SMT.
This is highlighted in figure~\ref{fig::mag_scan_density_B_SMT} (a,c,e) with a comparison of the profiles of $\langle n \rangle$ and $\langle |B| \rangle$, for $B_0 \in \{100, 300, 500\}$ in the SMT configuration. The profile of $B$ shifts to the HFS and follows the mean plasma density.

The shape and polarization of the poloidal component ($B_X, B_Z$) are also shown in figure~\ref{fig::mag_scan_density_B_SMT} (b,d,f). They clearly reveal a mode $m=+1$ and show that the polarization pattern follows the profiles of $\langle n \rangle$, as was observed in all the measurements presented in this work.
This behavior overall confirms the observations of figure~\ref{fig::comparison_helicon_density_Ar} (a,b,c) (PT configuration, $B_0=780$~G), and figure~\ref{fig::power_scan_n_Vf_helicon} (a,c,d) (PT configuration, $B_0=500$~G). In PT, while the polarization pattern of ($B_X, B_Z$) follows the density profile, the amplitude $\langle |B| \rangle$ does not.
In SMT, the addition of a vertical field for $B_0$ seems to link the helicon mode amplitude to the density profile.

Lastly, a global comparison of the plasma density and helicon mode amplitude is provided in figure~\ref{fig::mag_scan_density_B_SMT} (g), where the integrated profiles $\langle n\rangle_{t,X}$ and $\langle |B| \rangle_{t,X}$ are plotted for the entire dataset of these magnetic field scans. A global decrease of the helicon mode amplitude is observed when $B_0$ is raised. 
This trend cannot be interpreted from the linear perturbation model, which provides the dispersion relation and a link between $k$, $\omega$ as a function of the plasma density and $B_0$, but where the amplitude of $B$ does not appear in the equation.
The simple model predicts that when $k_r \gg k_z$, which is a reasonable assumption in TORPEX, the helicon wave damping fulfills $\gamma \propto 1/B_0 $ \citep{Chen_1991}. The results of figure~\ref{fig::mag_scan_density_B_SMT} (g) are in contradiction with this prediction, which shows the limits of the simple cylindrical model in interpreting TORPEX results.
The mechanisms behind the decrease of $|B|$ with $B_0$ are still not understood and require more in-depth investigations that are out of the scope of this article.

\section{Conclusions}
\label{sec::conclusion}

In this work, we have presented the first extensive experimental characterization of toroidal helicon waves in the low-temperature plasma device TORPEX. Using a birdcage resonant helicon antenna operating at 13.56~MHz, helicon waves are successfully launched in a pre-existing magnetron-generated plasma. The antenna is also used as a plasma source, over a wide range of parameters including the antenna power, magnetic field strength, and configuration, in argon and hydrogen.

A detailed identification of the dominant helicon mode is performed using B-dot probe measurements. By comparing the measured magnetic fluctuation profiles, local polarization, and eccentricity to a simplified analytical model assuming a uniform cylindrical plasma, we identify a dominant $m=+1$ helicon mode across all investigated regimes. Despite the simplifying assumptions of the model, a remarkable qualitative agreement with experimental measurements is observed, demonstrating the robustness of the $m=+1$ mode structure even in a toroidal geometry and in the presence of plasma non-uniformities.

Launching helicon waves at low power on a background magnetron-generated plasma allows us to investigate wave propagation properties with limited modification of the plasma equilibrium. In a purely toroidal magnetic field, the helicon wave exhibits a strong preferential propagation along the direction of the magnetic field and significant damping after a full toroidal turn. In contrast, the addition of a small vertical magnetic field in the SMT configuration enables nearly symmetric wave amplitudes on both sides of the antenna, indicating efficient toroidal propagation and strongly reduced damping. In both configurations, the local polarization of the helicon wave is found to closely follow the plasma density profile, while the wave amplitude showed little correlation with density.

When operating the helicon antenna as a plasma source, plasma generation up to densities of the order of $10^{17}$~m$^{-3}$ is achieved. In many laboratory devices using a helicon plasma source, an increased plasma generation performance is associated with helicon waves.
We show here that the plasma density in TORPEX scales linearly with the antenna power, and that beyond a threshold power of $P_H = 600$~W, this trend continues, irrespective of the helicon amplitude that instead saturates. This saturation coincides with a sharp increase in low-frequency density fluctuations and turbulent particle transport, suggesting a nonlinear coupling between the helicon wave and low-frequency instabilities. The strong correlation observed between helicon amplitude, low-frequency fluctuations, and turbulent transport point to energy transfer mechanisms from the RF helicon wave to low-frequency plasma dynamics.

The impact of the magnetic field $B_0$ on the plasma and helicon mode has also been explored. It is found that although the plasma equilibrium is substantially modified by changing $B_0$, the integrated plasma density, hence global plasma generation, remains unaffected. Increasing values of $B_0$ also tend to lower the helicon wave amplitude. These observations may motivate the exploration of helicon plasma wall-conditioning in tokamaks at lower magnetic field, where a higher RF field amplitude could be beneficial, without altering the plasma production.
 
Overall, and with its helicon antenna, TORPEX is a unique platform for the controlled study of toroidal helicon waves and their interaction with plasma turbulence.
Following the present study, future work will focus on quantitative measurements of toroidal wave numbers, improved modeling including toroidicity and density gradients, and a deeper investigation of the nonlinear coupling mechanisms between helicon waves and low-frequency plasma instabilities.

TORPEX is also equipped with a fast-ion source and detectors, as well as a central solenoïd. Future studies might take advantage of this equipment to explore the impact of helicon wave on fast ion propagation and diffusion, and on run-away electron mitigation.

\newpage

\appendix 

\section{Cylindrical helicon waves in uniform plasma density}
\label{appendix::simple_modeling}



\subsection{Whislter waves in cylindrical coordinates}

To plot the analytical solution of helicon waves in a simplified situation of a cylindrical geometry and uniform density, we follow the derivation of~\citep{Chen_1991}. We use a linear perturbation treatment in a cylindrical coordinate system $(r, \theta, z)$, assuming the waveform component evolves as $\underline{A}(r,\theta,z,t) = A(r) e^{i(m\theta + k_z z - \omega t)}$, with $m$ and $k_z$ the azimuthal and axial wavenumbers respectively. 
We solve the equation of motion for electrons coupled to Maxwell's equations, under the assumptions of uniform density, no collisions with neutrals, no displacement current, and no parallel electric field, which yields $ \bm{\nabla} \times \bm{B} = \alpha \bm{B}$, where $\alpha = \frac{\omega}{k_z} \frac{\mu_0 e n}{B_0}$. 
Applying $\nabla\times$ to equation~\ref{eq::B=alphaB} further leads to:
\begin{align}
    \partial_z^2 B_z + \frac{1}{r}\partial_z B_z +  \left(k_r^2 - \frac{m^2}{r^2}\right) B_z  = 0
\end{align}

where we define $k_r^2 = \alpha^2 - k_z^2$, and therefore denote $k=\alpha$. Notice that $k_r$ is not in the initial waveform evolution, but is defined a posteriori. It can therefore be thought rather as an effective radial wavenumber, providing radial shape of $B$, than linked to a precise radial wavelength.
The resolution gives $B_z$, and the components $(B_r, B_{\theta})$ are then computed from Eq.\eqref{eq::B=alphaB}. After simple algebra we have for a given azimuthal number $m$:

\begin{align}
    \begin{cases}
    B_r & = C [(k+k_z)J_{m-1}(k_r r) + (k-k_z)J_{m+1}(k_r r)] \cos(\omega t - k_z z) \\
    B_{\theta} & = -C [(k+k_z)J_{m-1}(k_r r) - (k-k_z)J_{m+1}(k_r r)] \sin(\omega t - k_z z) \\ 
    B_z & = 2C k_rJ_{m}(k_r r)\sin(\omega t - k_z z)
    \end{cases}
\end{align}
\label{eq::helicon_analytical_solution}

with $J_m$ the Bessel function of order $m$, and $C$ an arbitrary constant.
The electric components are given by  $E_{r} = \omega/kB_{\theta}$, $E_{\theta} = -\omega/kB_{r}$, $E_z = 0$.

\subsection{Boundary condition and helicon dispersion relation}

Let us consider cylindrical boundary conditions. A conductor would imply $E_{\theta}=0$, while a dielectric would impose no azimuthal current, so $j_{r} = (\bm{\nabla} \times \bm{B}) .\bm{e}_{r} / \mu_0  = \alpha/mu_0 B_{r} = 0$. In both cases this yields $B_r = 0$. From equation~\ref{eq::helicon_analytical_solution}, this means that the values of $x = k_r a$ are given by the roots of the equation:
\begin{equation}
    m \sqrt{k_z^2+k_r^2} J_m(k_r a) + k_z k_r a J_m'(k_r a) = 0 
\end{equation}
\label{eq::helicon_boundary_condition}

\begin{figure}
	\centering
	\begin{minipage}{.55\linewidth}
        \includegraphics[width = \columnwidth, trim={0in 0in 0in 0in},clip]{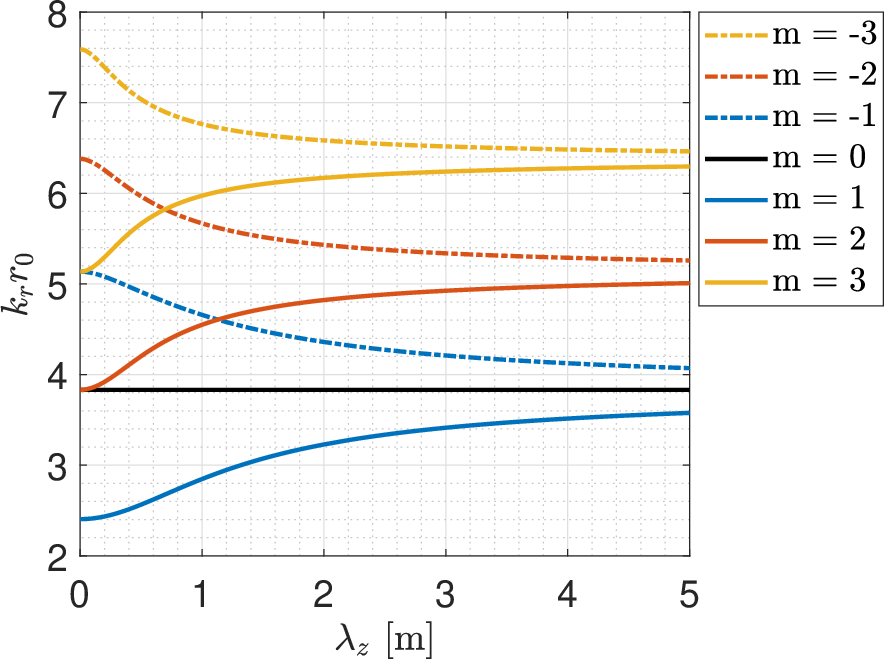}
    \end{minipage}
    \begin{minipage}{.38\linewidth}
        \begin{center}
        \begin{tabular}{c | c}
            $m$ & $k_r r_0$ (for $\lambda_z = 1.5$~m)  \\ 
            \hline
            -1 & 4.48 \\  
            0 & 3.83 \\
            +1 & 3.07 \\
            +2 & 5.51
        \end{tabular}
        \end{center}
    \end{minipage}
    \caption{Left: solutions of the equation setting the boundary condition $B_{\theta}=0$ as a function of the parallel wavelength $\lambda_z$, for different modes $m$. Right: values of $k_r r_0$ used to compute the helicon mode poloidal shape, considering $\lambda_z = 1.5$~m.}
    \label{fig::solution_kr_r0_boundary_condition}
\end{figure}

For a given radius of the cylinder $a$, we solve this equation for $x=k_r a$. This boundary condition therefore sets the value of the radial wavelength $k_r$. 
Figure~\ref{fig::solution_kr_r0_boundary_condition} shows the resulting values of $k_r a$ for the modes $m\in \{ -3; -2; -1; 0; 1; 2; 3\}$, as a function of the parallel wavelength $\lambda_z = \frac{2\pi}{k_z}$.

\subsection{Poloidal shape of the helicon modes}

The resulting magnetic fields for $m={-1, 0, 2}$ are plotted in figure~\ref{fig::helicon_Bpol_m=-1,0,2}. These helicon modes are computed assuming a cylindrical system $(r, \theta, z)$, with $z$ along the external magnetic field $B_0$.
We plot the results along the coordinates defined as $(x,y,z) \equiv (-r, z, \theta)$, to make the comparison easier with the simple toroidal coordinate system of TORPEX $(X, \phi, Z)$, with $\phi$ along $B_0$.

\begin{figure}
    \centering
    \includegraphics[width = 0.98\columnwidth, trim={0in 0in 0in 0in},clip]{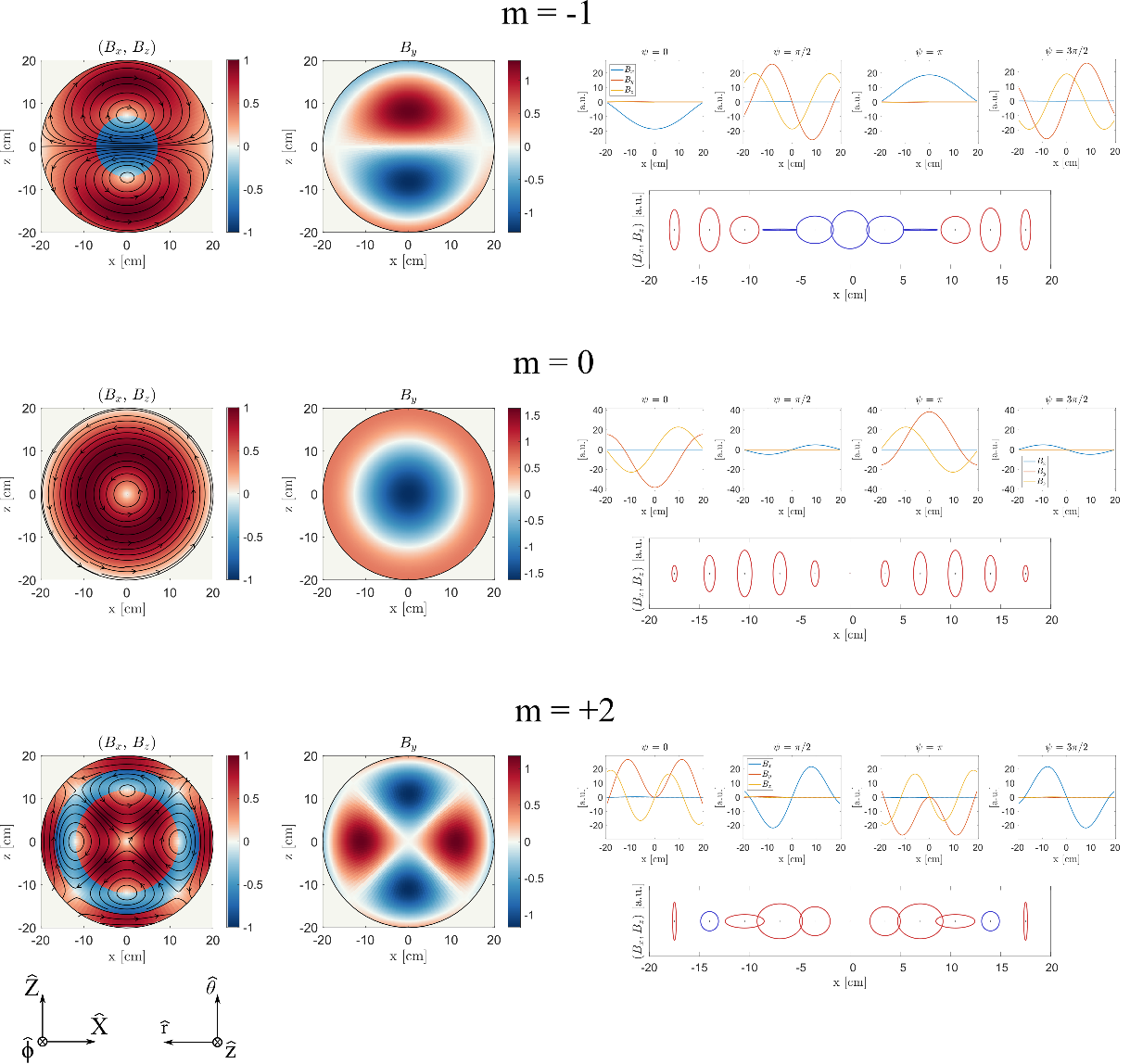}
    \caption{Modeling of helicon modes $m=-1$ (top), $m=0$ (middle) and $m=2$ (bottom).}
    \label{fig::helicon_Bpol_m=-1,0,2}
\end{figure}

\newpage

\section{Magnetron-generated plasma: additional data}
\label{appendix::instant_profiles}

\begin{figure}
    \centering
    \includegraphics[width = 0.99\columnwidth, trim={0in 0in 0in 0in},clip]{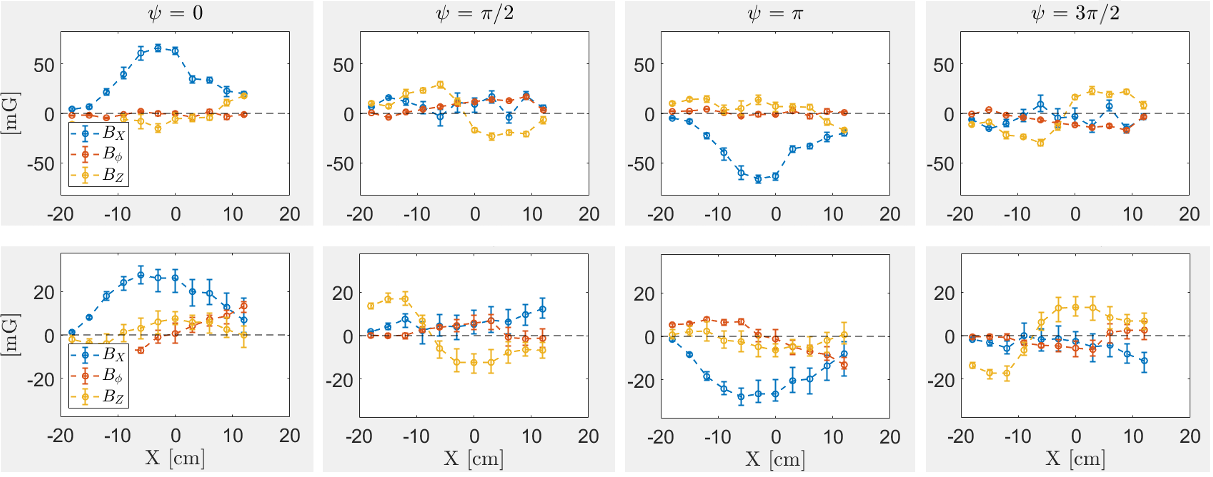}
    \caption{Argon, $P_{MAG} = 600$~W, $P_{H} = 200$~W, in PT (up) and SMT (bottom) configurations. These profiles are obtained by averaging the instantaneous profiles at these phases, over the entire duration of the 1~ms measurements.}
    \label{fig::instantaneous_profile_magnetron_Ar}
\end{figure}

\begin{figure}
    \centering
    \includegraphics[width = 0.99\columnwidth, trim={0in 0in 0in 0in},clip]{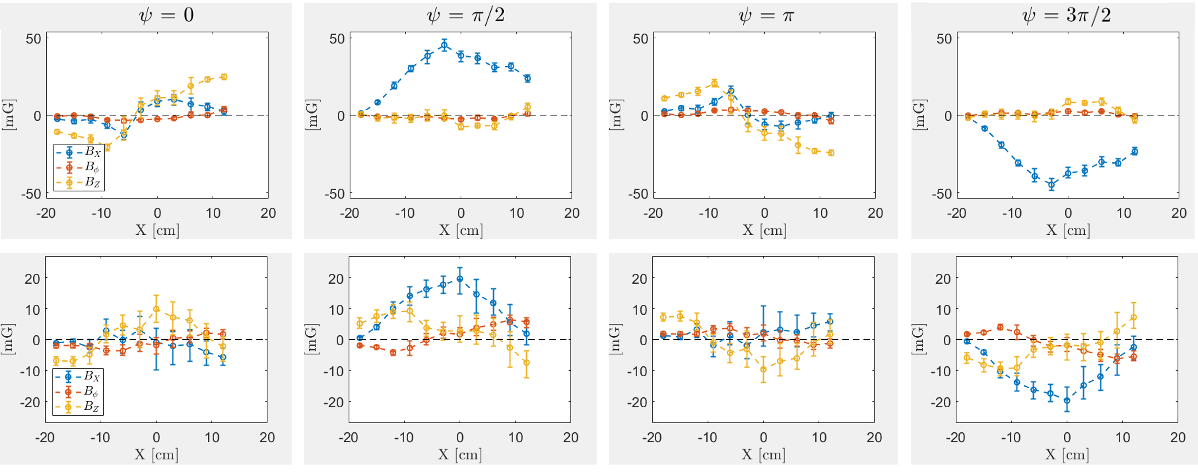}
    \caption{Hydrogen, $P_{MAG} = 600$~W, $P_{H} = 200$~W, in PT (up) and SMT (bottom) configurations. These profiles are obtained by averaging the instantaneous profiles at these phases, over the entire duration of the 1~ms measurements.}
    \label{fig::instantaneous_profile_magnetron_H2}
\end{figure}

\begin{figure}
    \centering
    \includegraphics[width = 0.9\columnwidth, trim={0in 0in 0in 0in},clip]{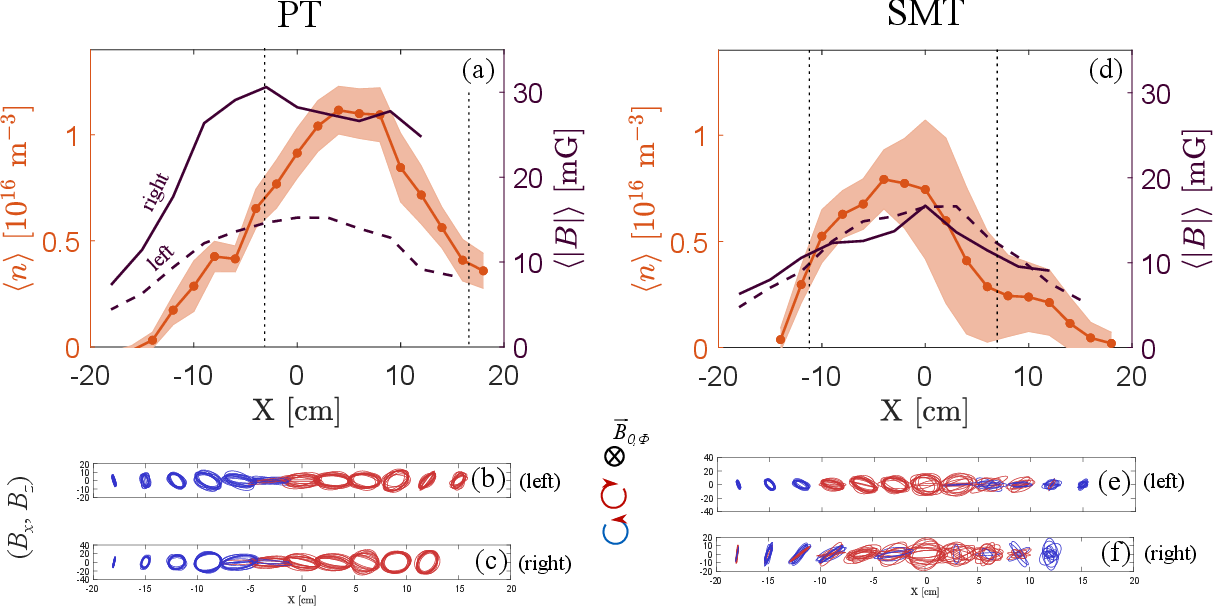}
    \caption{Same set of measurements as presented in figure~\ref{fig::comparison_helicon_density_Ar}, but in \textbf{hydrogen} plasmas. Measurements performed along $X$, at $Z=0$, for $P_{MAG} = 600$~W and $P_{H} = 200$~W.}
    \label{fig::comparison_helicon_density_H2}
\end{figure}

\begin{figure}
    \centering
    \includegraphics[width = 0.7\columnwidth, trim={0in 0in 0in 0in},clip]{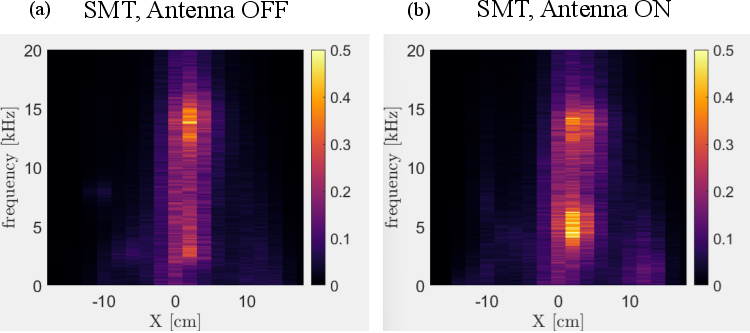}
    \caption{Spectra of the density fluctuations as a function of the position $X \in [-14:2:18]$~cm at ($Z=0$, $\phi \simeq + 80 \, ^{\circ}$). Measurements were performed in \textbf{hydrogen} plasma in the SMT configuration with $B_0=780$~G, for $P_{MAG} = 600$~W, and with the helicon antenna power (a) turned off and (b) turned on at $P_{H} = 200$~W.}
    \label{fig::spectra_density_fluct_H2}
\end{figure}

\begin{figure}
    \centering
    \includegraphics[width = 0.98\columnwidth, trim={0in 0in 0in 0in},clip]{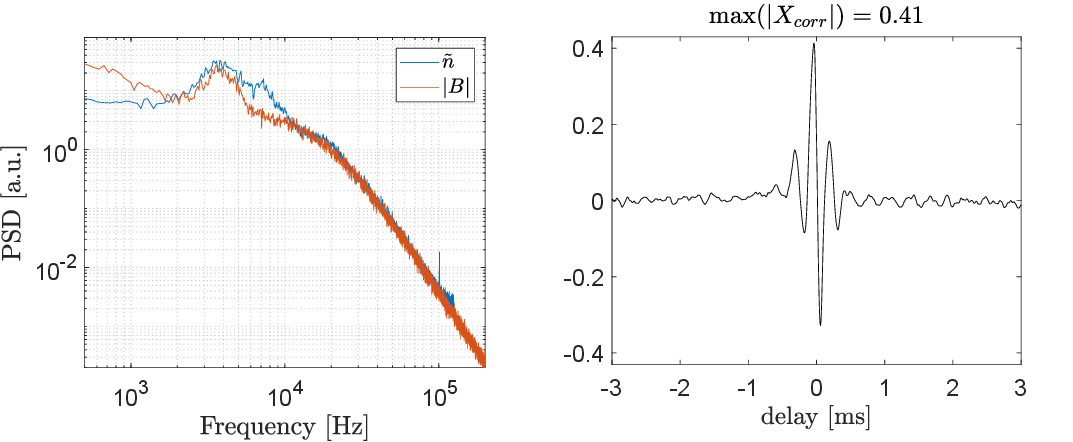}
    \caption{Left: spectra of the density fluctuations $\tilde{n}$ and of magnetic fluctuation amplitude $|B|$. Right: cross-correlation between $\tilde{n}$ and $|B|$. Experiments performed with $P_{MAG} = 600$~W, $P_{MAG} = 200$~W, in hydrogen plasma and SMT configuration with $B_0 = 785$~G.}
    \label{fig::simultaneous_meas_B_n}
\end{figure}

Additional experimental measurements are provided in this subsection, to support the finding disclosed in the main body of this article. Figure~\ref{fig::instantaneous_profile_magnetron_Ar} and~\ref{fig::instantaneous_profile_magnetron_H2} show, in argon and in hydrogen plasmas respectively, the 1D profiles of B-dot measurements of $(\tilde{B}_X, \tilde{B}_{\phi}, \tilde{B}_Z)$ performed in PT and SMT configurations. 
A detailed comparison between helicon waves measured on both sides of the antenna and the plasma density, in both PT and SMT configuration, is provided in figure~\ref{fig::comparison_helicon_density_H2}. Figure~\ref{fig::spectra_density_fluct_H2} shows, in hydrogen plasma, the enhancement of LF waves when the helicon antenna is turned on.
Figure~\ref{fig::simultaneous_meas_B_n} finally shows that density fluctuations and the helicon fluctuations in the kHz range are strongly correlated.

\newpage

\section*{Aknowledgments}

This work has been carried out within the framework of the EUROfusion Consortium, partially funded by the European Union via the Euratom Research and Training Programme (Grant Agreement No 101052200 — EUROfusion).
The Swiss contribution to this work has been funded in part by the Swiss State Secretariat for Education, Research and Innovation (SERI).
Views and opinions expressed are however those of the author(s) only and do not necessarily reflect those of the European Union, the European Commission or SERI. 
Neither the European Union nor the European Commission nor SERI can be held responsible for them.

\bibliographystyle{jpp}
\bibliography{mybib_helicon_torpex_JPP}

\end{document}